\documentclass{sig-alternate}

\usepackage{url}
\usepackage{array}
\usepackage{multirow}
\usepackage[labelformat=simple]{subcaption}
\usepackage[linesnumbered,longend,ruled,vlined]{algorithm2e}
\SetAlFnt{\footnotesize}
\SetAlCapFnt{\footnotesize}
\SetAlCapNameFnt{\footnotesize}
\usepackage{fixltx2e}
\usepackage{amssymb}
\usepackage{amsmath}
\usepackage{cite}
\usepackage{float}
\usepackage{tabularx,colortbl}
\usepackage{tikz}
\usepackage{enumitem}

\let\oldnl\nl
\newcommand{\nonl}{\renewcommand{\nl}{\let\nl\oldnl}}
\newcolumntype{?}{!{\vrule width 0.5pt}}

\newtheorem{definition}{Definition}
\newtheorem{example}{Example}

\newtheorem{property}{Property}

\newcommand{\RNum}[1]{\uppercase\expandafter{\romannumeral #1\relax}}

\begin{document}





%

\title{ROSIE: Runtime Optimization of SPARQL Queries Using Incremental Evaluation}

 \numberofauthors{3} 
 \author{
       \alignauthor Lei Gai\\      
      \email{lei.gai@pku.edu.cn}
       \alignauthor Wei Chen\\     
       \email{pekingchenwei@pku.edu.cn}
       \alignauthor Tengjiao Wang\\    
       \email{tjwang@pku.edu.cn}
 \end{tabular}\newline\begin{tabular}{c}
       \affaddr{School of Electrical Engineering and Computer Science}  \\
       \affaddr{Peking University. Beijing, P.R.China}   \\
}

\maketitle
\begin{abstract}

Relational databases are wildly adopted in RDF  (Resource Description Framework) data management. For efficient SPARQL query evaluation, the legacy  query optimizer needs reconsiderations. One vital problem is how to tackle the suboptimal query plan caused by error-prone cardinality estimation. Consider the schema-free nature of RDF data and the \textsc{Join}-intensive characteristic of SPARQL query, determine an optimal execution order before the query actually evaluated is costly or even infeasible, especially for complex queries on large-scale data. In this paper, we propose ROSIE, a \underline{R}untime \underline{O}ptimization  framework that iteratively re-optimize \underline{S}PARQL query plan according to the actual cardinality derived from \underline{I}ncremental partial query \underline{E}valuation. By introducing an approach for heuristic-based plan generation, as well as a mechanism to detect cardinality estimation error at runtime, ROSIE relieves the problem of biased cardinality propagation in an efficient way, and thus is more resilient to complex query evaluation. Extensive experiments on real and benchmark data show that compared to the state-of-the-arts, ROSIE consistently outperformed on complex queries by orders of magnitude. 

\end{abstract}


%
%



%
%
\section{Introduction} \label{sec-introduction}
The RDF\footnote{\scriptsize{\url{https://www.w3.org/RDF/}}} data model and SPARQL\footnote{\scriptsize{\url{http://www.w3.org/TR/sparql11-query}}} query language are W3C (World Wide Web Consortium) recommended standards for representing and accessing the ever-increasing linked resources, e.g., DBpedia\cite{DBpedia}, Freebase\cite{freebase}, UniProt\cite{uniprot}, etc. In practice, RDF data are physically managed by infrastructures, which can be dedicated RDF stores developed from scratch, e.g., \cite{Weiss:2008:HSI:1453856.1453965,rdf3x2010,triplebit-vldb2013}, or using legacy relational databases (RDB) as back-end,e.g.,\cite{virtuoso-bitmap,Abadi:2007:SSW:1325851.1325900,Broekstra:2002:SGA:646996.711426,Bornea:2013:BER:2463676.2463718}. RDB-backed systems can benefit greatly from 40+ years of advances in database domain, and have been an active research topic in both research literatures and industries. However, in the context of RDF/SPARQL, query optimization strategies directly borrowed from relational optimizer are not universally applicable\cite{Harth:2012:DTL:2213836.2213909}. This is especially true for complex queries on large-scale data. In general, there are two noteworthy problems that need to be carefully reconsidered.

First, due to the schema-free nature of RDF, cardinality estimation is error-prone, and sometimes even biased by orders of magnitude. In a traditional query optimizer, cardinality estimation is the dominating component that affects optimal query plan selection\cite{cardproblem}. It is canonically based on the containment and independence assumption. To tackle correlations in data, the most widely adopted and cost-effective method is by maintaining pre-build statistics as multi-dimensional histograms\cite{Ioannidis:2003:HH:1315451.1315455}. For RDF data, it is commonly represented as a collection of \textit{triple}s in the form of three elements, $\langle$\textit{Subject, Predicate, Object}$\rangle$, (or $\langle$S,P,O$\rangle$ for brevity). Even though comprehensive statistics of all joinable combinations can exactly capture correlations in such data, but they tend to be exponentially huge, as each triple may has correlations with others. Existing researches proposed RDF-specific statistical synopses, such as permutation indexes\cite{Weiss:2008:HSI:1453856.1453965}, possible join path\cite{rdf3x2010} or \textit{Characteristic Sets}\cite{Neumann:2011:CSA:2004686.2005604}, in order to give better estimates than always assuming independence. But they are still insufficient for more complex queries. Besides, managing and accessing these statistics needs a redesign of the underlying RDF storage. This is not well suited for RDB-backed systems. 

Second, consider the SPARQL's \textsc{Join}-intensive characteristic, the propagation of cardinality estimate errors has a more profound effect on query performance. A SPARQL query is typically specified as a combination of \textit{triple pattern}s\footnote{\scriptsize{https://www.w3.org/TR/sparql11-query/\#QSynTriples}}  (TP) in a fine-grained manner. The query plan (a.k.a., execution order) consists of significantly more \textsc{Join}s caused by each single access of a TP. Earlier work has shown that errors in cardinality estimate may increase exponentially with the number of \textsc{Join}s\cite{Ioannidis:1993:OHL:169725.169708}. Although carefully cultivated storage, e.g., property tables\cite{Chong:2005:ESR:1083592.1083734,wilkinson06} or vertically partitioned columns\cite{Abadi:2007:SSW:1325851.1325900}, can alleviate this problem by reducing the number of \textsc{Join}s, they did not solve the fundamental problem of cardinality estimate error. Furthermore, in real-world queries, syntactic constructs, such as OPTIONAL,UNION and FILTER etc, are extensively used\cite{Picalausa:2011:RSQ:1999299.1999306}. These complicated the problem.

To tackle these problems, runtime optimization approaches \cite{Cole:1994:ODQ:191839.191872, Graefe:1989:DQE:67544.66960, Bruno:2002:ESQ:564691.564722,AbdelKader:2009:RRO:1559845.1559910, DBLP:conf/btw/0001G13} are promising.  This kind of approaches incrementally materialize the intermediate results during query execution time, gather the accurate cardinalities and eliminate errors at given steps, then use these feedbacks to dynamically adjust the query plan. In this case, cost-efficient heuristics-based methods \cite{DBLP:conf/edbt/Gubichev014, Stocker:2008:SBG:1367497.1367578,Neumann:2011:CSA:2004686.2005604} are feasible for query plan generation. Such plan is not guaranteed to be optimal, but can be dynamically optimized at runtime. Nonetheless, consider the pervasive existence of interactions between \textsc{Join} predicates in a SPRAQL query, directly applying above methods may results in a materialization at each query execution step. This can greatly down-grades the SPARQL query performance as materialization is costly. Therefore, two challenges arise: (\textit{i}) how to generate an optimized TP execution order that needs as fewer materialization steps as possible; (\textit{ii}) How to determine at which step an incremental materialization is needed during runtime optimization.
 
In this paper, we propose ROSIE, a framework for \underline{R}untime \underline{O}ptimization of \underline{S}PARQL query using \underline{I}ncremental \underline{E}valuation. The basic idea of ROSIE is based on that, as an optimal execution order is hard to be pre-determined in a cost-effective way, runtime evaluation of partial query at a given step can eliminate the cardinality estimation error that may propagate to the following steps, and materialized intermediate results can be used to guide the optimal selection of the remaining execution order. It starts with an initial TP execution order, named \textit{Candidate Sequence} ($\mathcal{CS}$), then sequentially determines the cardinality estimation error at each step in $\mathcal{CS}$. Incremental evaluation of partial query is triggered once ROSIE detects that current step may lead to a suboptimal execution. ROSIE follows a RDB-backed infrastructure. This guarantees the efficiency of $\mathcal{CS}$ evaluation and intermediate results materialization. The following techniques tackle the above challenges brought by adopting runtime optimization.

For the first challenge, we propose a heuristic-based method for $\mathcal{CS}$ generating. Different from existing works that mainly focus on TP \textsc{Join}s, we introduce a query graph, named as \textit{Query Relation Graph} (QRG), to model the query structure of more versatile semantics in SPARQL that includes UNION, FILTER and OPTIONAL, etc. To generate a $\mathcal{CS}$ that optimally guides the query execution, we present a greedy algorithm to traversal QRG, which synthesizes the semantical equivalence of SPARQL operators and preliminary statistics of RDF data. This is detailed in \textit{Section} \ref{sec:initial}.

For the second challenge, we devise a mechanism for possible suboptimal steps detection. It utilizes the propagation of cardinality estimation errors in a $\mathcal{CS}$. We argue that, for an optimal $\mathcal{CS}$, the estimation error bound is propagated in a non-incremental fashion. Consider that the error estimate needs not to be as accurate as the cardinality estimate, the general statistics maintained by a RDB  are useful in determining a feasible error bound. This makes ROSIE more practical as it does not need to revise the implementation of the underlying RDB. This is detailed in \textit{Section} \ref{sec:optimization}.

To summarize, our contributions in this paper are three-fold: (\textit{i}) We introduce a radical new runtime optimization framework, named ROISE. It uses incremental partial query evaluation to dynamically optimize the query execution, and is characterized by supporting efficient evaluation of more expressive SPARQL queries; (\textit{ii}) To boost the performance of ROSIE, we propose a heuristics-based approach for generating an optimized TP execution order, and devise a mechanism that leverage error bounds estimate for possible suboptimal steps detection; (\textit{iii}) Through extensive experiments on benchmark and real datasets, we show the correctness and efficiency of ROSIE. For complex queries, ROSIE yields performances that are 1.5-20 times better than the state-of-the-arts. 

The rest of the paper is organized as follows: \textit{Section} \ref{sec:related} reviews the most related works. \textit{Section} \ref{sec:preliminaries} introduces the principal concepts used. \textit{Section} \ref{sec:overview} gives an overview of ROSIE. \textit{Section} \ref{sec:initial} details the heuristic-based method for $\mathcal{CS}$ planning, and the mechanism for determining the error bound propagation is introduced in \textit{Section} \ref{sec:optimization}. \textit{Section} \ref{sec:experiments} shows the experimental results, and \textit{Section} \ref{sec:conclusions} concluded.

%
%
\section{Related Work} \label{sec:related}
Cardinality estimation is still an active topic, as it tends to be more error-prone among all the costs involved in query optimization\cite{Chaudhuri:1998:OQO:275487.275492}. Existing literatures can be roughly classified as \textit{static} approaches or \textit{runtime} approaches. 

\textit{Static} approaches rely on off-line maintained statistics, such as histograms (e.g., \cite{Ioannidis:2003:HH:1315451.1315455}) or samples (e.g., \cite{Lipton:1990:PSE:93597.93611}), etc, to determine a query plan that is unalterable throughout the query execution. In the context of RDF/SPARQL, maintaining all possible statistics is unrealistic. Statistics of permutation indexes \cite{Weiss:2008:HSI:1453856.1453965}, or even more complicated statistics, such as \textit{Characteristic Sets}, \cite{Neumann:2011:CSA:2004686.2005604} are proposed. They only delay the step where errors propagated, thus are less effective for complicated queries with more \textsc{Join}s. Heuristics-based approaches, e.g., using greedy strategy\cite{Stocker:2008:SBG:1367497.1367578} or query structure\cite{Tsialiamanis:2012:HQO:2247596.2247635}, are effective in pruning the plan search space, and are extensively adopted as subsidiary mechanisms\cite{rdf3x2010,triplebit-vldb2013,DBLP:conf/edbt/Gubichev014}. But they are not guaranteed to be optimal.

Other than attempting to pre-determine an optimal query plan, \textit{Runtime} approaches \cite{Cole:1994:ODQ:191839.191872, Graefe:1989:DQE:67544.66960, Bruno:2002:ESQ:564691.564722,AbdelKader:2009:RRO:1559845.1559910, DBLP:conf/btw/0001G13, Kotoulas:2012:RRO:2426516.2426533} optimize query plan at runtime when the actual cardinality turns available. There are few works of \textit{Runtime} optimization for RDF/SPARQL. The most relevant works in literatures are \cite{AbdelKader:2009:RRO:1559845.1559910}, \cite{Kotoulas:2012:RRO:2426516.2426533} and \cite{DBLP:conf/btw/0001G13}. \cite{AbdelKader:2009:RRO:1559845.1559910} materialized intermediate results at each step, which seems to be an overkill. \cite{Kotoulas:2012:RRO:2426516.2426533} proposed a similar framework as ROSIE, but is based on \cite{AbdelKader:2009:RRO:1559845.1559910} and focused on dealing with skews in a distributed scenario. \cite{DBLP:conf/btw/0001G13} is the most similar to ROSIE, as they all devise mechanism for detecting error propagation, which are used to selectively choose plan fragments to be executed. But the heuristics adopted in \cite{DBLP:conf/btw/0001G13} generally work well in RDBs, they need to be reconsidered for SPARQL queries.

%
%
\section{Preliminaries} \label{sec:preliminaries}
\subsection{RDF and SPARQL} \label{subsec-sparql-query}
In RDF data model, a real world \textit{resource} is identified by an unique URI (Uniform Resource Identifier) string, and each of its properties or linkages with other \textit{resource}s is depicted as a triple $t_i$ of three components, $t_i =\langle$S,P,O$\rangle$. A collection of such triples forms an \textit{RDF dataset} $\mathcal{D}= \bigcup^{|D|}_{i=1} t_i$.

Like SQL for relational database, SPARQL is the de facto query language for RDF.  Basically, a SPARQL query consists of a query form (i.e. \textsc{Select},\textsc{Ask}, \textsc{Construct} or \textsc{Describe}), a \textit{Basic Graph Pattern} (BGP) specified in a WHERE clause, and possibly solution modifiers (e.g. DISTINCT). The building block of BGPs are TPs. Each TP is expressed in a triple form, with a least one of S, O, P elements replaced by unbounded \textit{variable}s\footnote{\scriptsize{https://www.w3.org/TR/sparql11-query/\#QSynVariables}}. A BGP can be recursively defined as a TP, or a finite set of BGPs connected by SPARQL operators  (\textsc{And}, \textsc{Or}, \textsc{Opt}, and \textsc{Filter}) to build expressive queries\footnote{\scriptsize{Specifically, \textsc{And} connects two conjunctive BGPs, \textsc{Or} for BGPs connected by UNION keywords, \textsc{Opt} for OPTIONAL keywords, and \textsc{Filter} for FILTER conditions. }}. A TP \textsc{Join} is a \textsc{Join} (correspondence to \textsc{And} operator, analogous to equi-\textsc{Join}) between two TPs on a variable (or a join predicate in relational database literatures). A SPARQL query $\mathcal{Q}$ can be formalized as a quadruple, $\mathcal{Q} = \langle \mathbb{T}, \mathbb{O}, \mathbb{V}, \mathbb{S} \rangle$, where,
\begin{itemize}
    \item  $\mathbb{T}$ is the finite set of TPs in $\mathcal{Q}$,.
    \item  $\mathbb{O}$ is the set of SPARQL operators, For $\circledcirc \in \mathbb{O}$, $\circledcirc$ can be one of \textsc{And}, \textsc{Or}, \textsc{Opt} or \textsc{Filter}.
    \item  $\mathbb{V}$ is the \textit{variable correlations}, expressed as groups of TPs in $\mathbb{T}$ that share  common variables at given positions (S,O or P). 
    \item  $\mathbb{S}$ is the \textit{query semantics}, which states how TPs in $\mathbb{T}$ are connected by operators in $\mathbb{O}$. 
\end{itemize}
 
\begin{figure}
  \centering
  \includegraphics[width=2.3 in]{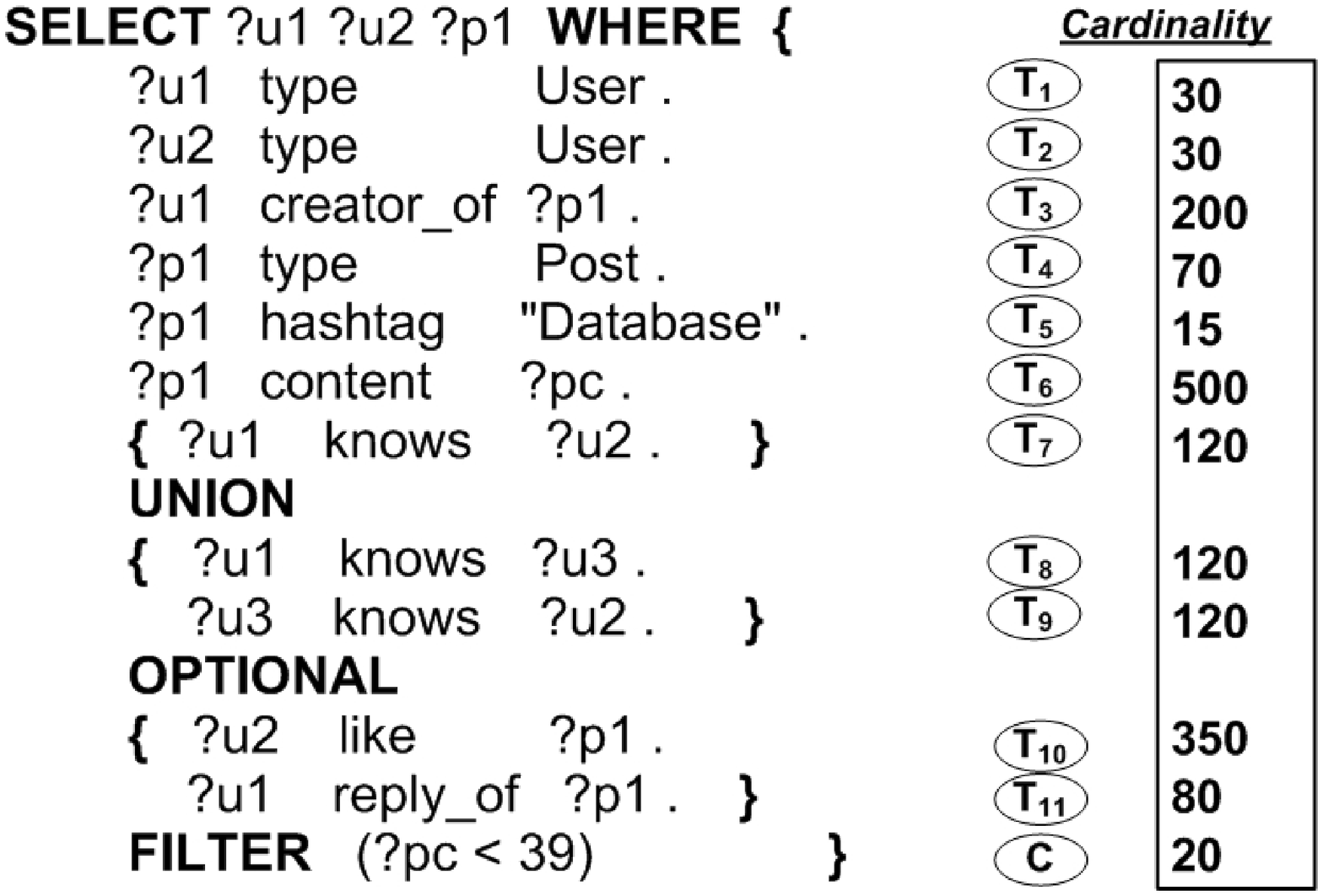}
  \captionsetup{belowskip=1.5pt,aboveskip=1.5pt}
  \caption{An example SPARQL query $\mathcal{Q}_e$.}
  \label{fig:sparql-query-example}
\end{figure}

\begin{example}
\small
Figure \ref{fig:sparql-query-example} shows an example of a \textsc{Select} query $\mathcal{Q}_e$. In its WHERE clause, 11 TPs and  a \textsc{Filter} constraint is defined \footnote{\scriptsize{Noteworthy that in order to make this example easier to read, we replaced the URIs with more readable names in $\mathcal{Q}_e$.}}. The formalization of $\mathcal{Q}_e$ is shown as Figure \ref{fig:query-formalization}. 
\end{example}

\begin{figure}
  \centering
  \includegraphics[width=1.8 in]{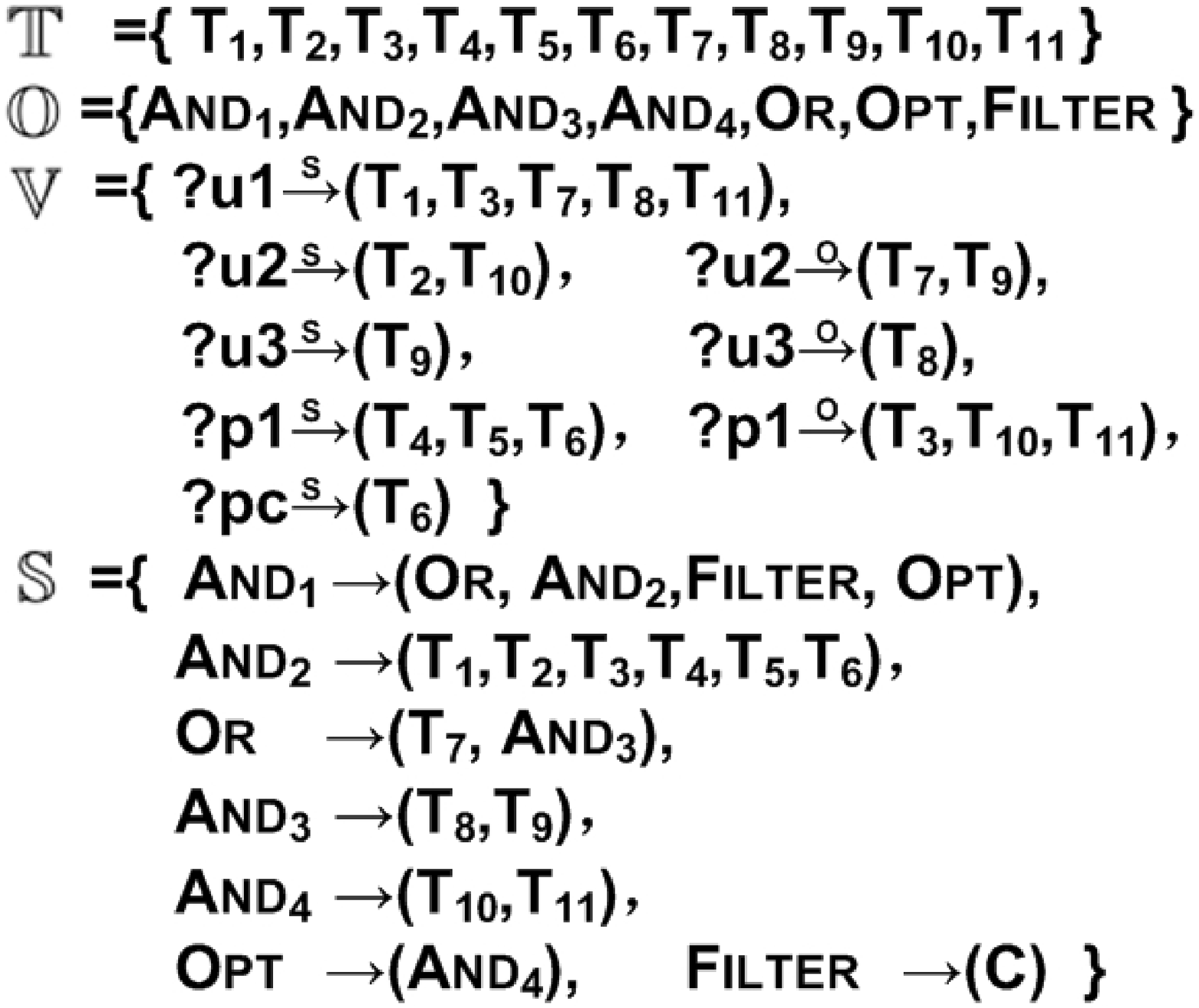}
  \captionsetup{belowskip=1.5pt,aboveskip=1.5pt}
  \caption{Formalization of $\mathcal{Q}_e = \langle \mathbb{T}, \mathbb{O}, \mathbb{V}, \mathbb{S} \rangle$.}
  \label{fig:query-formalization}
\end{figure}

\subsection{Candidate Sequence ($\mathcal{CS}$)}
Traditionally, a SPARQL query $\mathcal{Q}$ is evaluated in a TP-by-TP manner. The optimization goal is focused on determining an optimal operator tree of TPs that minimize the intermediate results size during query evaluation. In this paper, we formally define an representation of the operator tree, named as a \textit{Candidate Sequence} ($\mathcal{CS}$), as follows : 

\begin{definition}[Candidate Sequence]
Given a $\mathcal{Q} = \langle \mathbb{T}, \mathbb{O}, \mathbb{V}, \mathbb{S} \rangle$, a Candidate Sequence $\mathcal{CS}$ can be recursively defined as:
\begin{itemize}
    \item  $(T)$ is a $\mathcal{CS}$, where $T \in \mathbb{T}$.
    \item  If $\mathcal{CS}_m$, $\mathcal{CS}_n$ are $\mathcal{CS}$, for $\circledcirc \in \mathbb{O}$, $(\mathcal{CS}_m \circledcirc \mathcal{CS}_n)$ is also a $\mathcal{CS}$.
\end{itemize}
\end{definition}

The concept of $\mathcal{CS}$ are analogous to the query plan used in the relational \textit{cost-based query optimization} (CBO). A partial $\mathcal{CS}$ can be regarded as a \textit{step} in evaluation. As $\mathcal{CS}$ is expressed in the form of SPARQL algebra, like the relational equivalence query plans in CBO, a $\mathcal{Q}$ may have many \textit{equivalent} (denoted by $\rightleftharpoons$) $\mathcal{CS}$s\cite{Perez:2009:SCS:1567274.1567278}. Let $T_1$, $T_2$ and $T_3$ be TPs and $C$ is a constraint,  the following \textit{Properties} hold.

\begin{property}
\label{property-exchangable}(Exchangeable Region)
\small
The position of a TP in an \textsc{And}-only or an \textsc{Or}-only $\mathcal{CS}$ is exchangeable. 
\begin{enumerate}[topsep=0.5pt,itemsep=-0.5ex,partopsep=1ex,parsep=1ex]
\scriptsize
\item (($T_1$ \textsc{And} $T_2$) \textsc{And} $T_3$) $\rightleftharpoons$ (($T_1$ \textsc{And} $T_3$) \textsc{And} $T_2$).
\item (($T_1$ \textsc{Or} $T_2$) \textsc{Or} $T_3$) $\rightleftharpoons$ (($T_1$ \textsc{Or} $T_3$) \textsc{Or} $T_2$).
\end{enumerate}
\normalsize
\end{property} 
\textit{Property} \ref{property-exchangable} states that TPs connected by \textsc{And} or \textsc{Or} can freely exchange theirs positions in a $\mathcal{CS}$. It stands as both \textsc{And} and \textsc{Or} are commutative and associative. 

\begin{property} \label{property-commutative}(Distributable Region)
\scriptsize
\begin{enumerate}[topsep=0.5pt,itemsep=-0.5ex,partopsep=1ex,parsep=1ex]
\itemsep0em
\item ($T_1$ \textsc{And} ($T_2$ \textsc{Or} $T_3$)) $\rightleftharpoons$ (($T_1$ \textsc{And} $T_3$) \textsc{Or} ($T_1$ \textsc{And} $T_2$)).
\item ($T_1$ \textsc{Opt} ($T_2$ \textsc{Or} $T_3$)) $\rightleftharpoons$ (($T_1$ \textsc{Opt} $T_3$) \textsc{Or} ($T_1$ \textsc{Opt} $T_2$)).
\item (($T_1$ \textsc{Or} $T_2$) \textsc{Opt} $T_3$)) $\rightleftharpoons$ (($T_1$ \textsc{Opt} $T_3$) \textsc{Or} ($T_2$ \textsc{Opt} $T_3$)).
\item (($T_1$ \textsc{Or} $T_2$)\textsc{Filter} $C$) $\rightleftharpoons$ (($T_1$ \textsc{Filter} $C$)\textsc{Or}($T_2$ \textsc{Filter} $C$)).
\end{enumerate} 
\normalsize
\end{property} 
\textit{Property} \ref{property-commutative}  states that a latter TP can be distributed to its proceeding steps in $\mathcal{CS}$. Its correctness can be proven according to the SPARQL algebra equivalence stated in \cite{Perez:2009:SCS:1567274.1567278}.  \textit{Property} \ref{property-exchangable}  and \ref{property-commutative} are the foundations of equivalent $\mathcal{CS}$ generation used in this paper.

Recap that in CBO, cardinality estimation is performed at each step of a query plan to derive a cardinality of the contemporary result set (a.k.a. a cost). Given an RDF dataset $\mathcal{D}$, and a SPARQL query $\mathcal{Q} = \langle \mathbb{T}, \mathbb{O}, \mathbb{V}, \mathbb{S} \rangle$, we detail the problem of cardinality estimation in a $\mathcal{CS}$ from two aspects: the cardinality of TPs, and the cardinality of TP \textsc{Join}s.

The cardinality of a TP $T_i = \langle S,P,O \rangle \in \mathbb{T}$ is determined as $|T_i|=p(T_i) \times |\mathcal{D}|$, where $p(T_i)$ is the selectivity of $T_i$, and can be explained as the distribution of triples in $\mathcal{D}$ that match $T_i$. Clearly, $p(T_i) \in [0, 1]$. Under the independence assumption (i.e. S, P and O are mutually independent), $p(T_i)$ can be estimated by the formula $p(T_i) = p(S) \times  p(P) \times  p(O)$, where $p(\bullet)=1.0$ if element $\bullet \in \{S,P,O\}$ is  unbounded, and for bounded object, $p(\bullet)$ is approximated by means of classical single-value histograms in a RDB\cite{Stocker:2008:SBG:1367497.1367578}. In fact, independence assumption rarely holds in $\mathcal{D}$. 

The cardinality of a TP \textsc{Join} can be varied between 0 and the Cartesian product size. The join selectivity estimation is a further investigated problem for RDB with a well-defined schema. It general, the cardinality of \textsc{And} operator can be estimated as : 
\begin{equation}
|T_i \ \textsc{And}\ T_j | = p(T_i\ \textsc{And} \ T_j ) \times |T_i| \times |T_j|
\end{equation}
where $p(T_i\ \textsc{And} \ T_j )$  denotes the selectivity of \textsc{And}.  Under the containment assumption, $p(T_i\ \textsc{And} \ T_j ) = \frac{1}{max(p(T_i),p(T_j))}$\cite{Chaudhuri:1998:OQO:275487.275492}. This works well for foreign-key join. Consider the schema-free nature of $\mathcal{D}$, such estimate is error-prone. 

An incorrect single step in cardinality estimation potentially screws up whole execution plan. As maintaining comprehensive statistical informations in a RDB is costly and impractical, cardinality estimation errors are unavoidable. We define the cardinality estimation error as: 
\begin{equation} \label{equ-errordef}
\varepsilon_{\circledcirc}(T_i,T_j) = \frac{|T_i\circledcirc T_j|_P}{|T_i\circledcirc T_j|_E}
\end{equation}
, where $|.|_P$ denotes the real cardinality, and $|.|_E$ denotes the estimated cardinality. $\varepsilon < 1$ means an over-estimate, and $\varepsilon > 1$ means an under-estimate.  We further discuss the error estimation in \textit{Section} \ref{subsec-roebe}.

\begin{example}
\small
Consider $\mathcal{Q}_e$ in Figure \ref{fig:sparql-query-example}. We can assign a  $\mathcal{CS} = ((T_1\ \textsc{And}\ T_3) \textsc{And}\ T_6) $ for partial query evaluation. There $\mathcal{CS}_{Step 1} = (T_1\ \textsc{And}\ T_3)$, and $\mathcal{CS}_{Step 2} =(\mathcal{CS}_{Step 1}\ \textsc{And} \ T_6)$. Adopting traditional cardinality estimates method discussed above, we get $|\mathcal{CS}_{Step 1}|=30$, $|\mathcal{CS}_{Step 2}|=30$. Consider the semantic correlations exists in $\mathcal{D}$,  that is each Post is created by a User, and each Post has a content, the actual cardinalities are $|\mathcal{CS}_{Step 1}|=200$, $|\mathcal{CS}_{Step 2}|=500$. Thus, $\varepsilon_{\textsc{And}}(T_1,T_3)=\frac{200}{30}\approx6.7$, $\varepsilon_{\textsc{And}}(\mathcal{CS}_{Step 1},T_6)=\frac{500}{30}\approx16.7$. This states that current estimates are vastly under-estimated.
\end{example}

%
%

\section{Overview of ROSIE} \label{sec:overview}
The general idea of ROSIE boils down to determining the optimal $\mathcal{CS}$ dynamically during query evaluation. ROSIE is based on the rationale of pushing as much hard and time consuming operations into the RDB as possible. This can profit most from the existing highly optimized techniques, while using runtime optimization technique to alleviate the problems caused by RDF/SPARQL-specific features. The main algorithm is shown in Algorithm \ref{alg:overview}. It mainly consists of two phases.

\begin{algorithm}
\caption{\scriptsize{ROSIE algorithm.}}
\label{alg:overview}
\scriptsize{}
\SetKwInOut{Input}{input}
\SetKwInOut{Output}{output}
\Input{$\mathcal{Q}$, and $\mathcal{D}$ over a RDB.}
\Output{ $R$, evaluation results of $\mathcal{Q}$ on $\mathcal{D}$.}
\nonl \Begin{
$\mathcal{Q}$ parsing, and QRG $G_Q$ construction; \quad \ //\textbf{Section \ref{subsec-qrg}}\; \label{alg1:line:step1}
$\mathcal{CS} \Leftarrow $ CSConstruction($G_Q$);  \qquad   \qquad \qquad//\textbf{Section \ref{subsec-initcs}}\; \label{alg1:line:step2}
$T_1, T_2 \Leftarrow $ Get the first 2 TPs from $\mathcal{CS}$ \;
$\mathcal{CS}_{sub} \Leftarrow \{T_1\}$; $R_s \Leftarrow$ \{.\}; $\widehat{G_Q} \Leftarrow G_Q$; \quad \quad //\emph{Initialization} \; 
\While {having TPs not traversed in $Q_G$} {
    $\varepsilon \Leftarrow $ CardErrEsti($\mathcal{CS}_{sub}$, $T_i$, $T_{i+1}$);  \quad //\textbf{Section \ref{subsec-roebe}} \; \label{alg1:line:step4}
    \eIf {error bound exceeded} {
        $\mathcal{CS}_{sub} \mapsto $ SQL; \qquad \qquad //\emph{Convert $\mathcal{CS}_{sub}$ to SQL} \; \label{alg1:line:step5} 
        $R_i \Leftarrow$ SQL evaluation in RDB \; \label{alg1:line:step6}
        $\mathcal{CS}_{sub} \Leftarrow \{T_i\}$ ; $R_s \Leftarrow R_i$; $\widehat{G_Q} \Leftarrow $ update $\widehat{G_Q}$\; \label{alg1:line:step7}
        $\mathcal{CS} \Leftarrow $ CSConstruction($\widehat{G_Q}$); \qquad //\textbf{Section \ref{subsec-initcs}}\; \label{alg1:line:step8}
    }{
    $\mathcal{CS}_{sub} \Leftarrow \circledcirc, TP_i$   \; \label{alg1:line:step9}
    $T_i \Leftarrow $ Get next TP from $\mathcal{CS}$ \; \label{alg1:line:step10}
    }
}
\eIf{$\mathcal{CS}_{sub}$ is not null}{
	$\mathcal{CS}_{sub} \mapsto $ SQL; $R \Leftarrow$ SQL evaluation \; \label{alg1:line:step11} 
}{
	$R \Leftarrow R_s$ \;  \label{alg1:line:step11} \label{alg1:line:step12} 
}
}
\scriptsize{}
%
\end{algorithm}

The first phase analyzes the structure of a query and initializes a $\mathcal{CS}$. In essence, after a SPARQL query $\mathcal{Q}$ is submitted and parsed, a \textit{Query Relation Graph} (QRG) that models $\mathbb{V}$ and $\mathbb{S}$ in $\mathcal{Q}$ is constructed (\textit{Line} \ref{alg1:line:step1}, details in \textit{Section} \ref{subsec-qrg}). Then an initial $\mathcal{CS}$ is constructed by traversing the QRG following to a set of heuristics in a greedy manner (\textit{Line} \ref{alg1:line:step2}, details in \textit{Section} \ref{subsec-initcs}).

The second phase performs runtime optimization during query evaluation. It is achieved by iteratively determining an cardinality estimate error bound at each step of $\mathcal{CS}$. Once the augmentation of current TP tends to cause cardinality estimate unacceptably biased, the equivalent SQL of the partial $\mathcal{CS}$ that contains all ancestor TPs is generated and submitted to the underlying RDB for execution (\textit{Line} \ref{alg1:line:step4}-\ref{alg1:line:step8}, details in \textit{Section} \ref{subsec-roebe}). The materialized results of the partial $\mathcal{CS}$ evaluation, as well as data statistics, are managed by the underlying RDB, and are used in the next iteration to plan  $\mathcal{CS}$ for the remaining part of query at runtime (\textit{Line} \ref{alg1:line:step7}-\ref{alg1:line:step8}). Otherwise, the algorithm just proceed with the next TP in $\mathcal{CS}$ (\textit{Line} \ref{alg1:line:step9}-\ref{alg1:line:step10}). This phase iterates until all TPs have been processed, and returns the final results (\textit{Line} \ref{alg1:line:step11}-\ref{alg1:line:step12} ). 

In \textit{Algorithm} \ref{alg:overview}, to minimize the materialization costs, ROSIE focus on the optimization of two key operations: the $\mathcal{CS}$ planning (\textit{Line} \ref{alg1:line:step2}  and \ref{alg1:line:step8}) and the cardinality estimate error determination (\textit{Line} \ref{alg1:line:step4}). We details them in \textit{Section} \ref{sec:initial} and \ref{sec:optimization} respectively.

%
%
\section{Candidate Sequence Planning} \label{sec:initial}
In this section, we introduce a method to generate an optimized $\mathcal{CS}$. It models the query structures into a graph structure, named as \textit{Query Relation Graph} (QRG). We explain the formation of QRG in \textit{Section} \ref{subsec-qrg}, and devise a novel linear-time heuristics-based greedy algorithm for $\mathcal{CS}$ planning in \textit{Section} \ref{subsec-initcs}.

\subsection{Query Relation Graph (QRG)} \label{subsec-qrg}
As discussed in Section \ref{subsec-sparql-query}, a SPARQL query $\mathcal{Q}$ have its \textit{variable correlations} $\mathbb{V}$ and \textit{query semantics} $\mathbb{S}$. To well model $\mathcal{Q}$, we introduce the concept of Query Relation Graph (QRG). It can be formally defined as :
\begin{definition}[Query Relation Graph, QRG]
The QRG of a SPARQL query $\mathcal{Q}$ is a directed, vertex-weighted, edge-labeled graph, denoted as $G_\mathcal{Q}=(V,E,f_w,f_l)$, where: 
\begin{itemize}
\item Vertex set $V=V_{\RNum{1}} \cup V_{\RNum{2}} \cup V_{\RNum{3}} $. $v_\circledcirc \in V_{\RNum{1}}$ denotes an operator, $v_t \in V_{\RNum{2}}$ denotes a triple pattern, and $v_v \in V_{\RNum{3}}$ denotes an unbounded variable in $\mathcal{Q}$.
\item Edge set $E=E_{\RNum{1}} \cup E_{\RNum{2}} $. $e \in E_{\RNum{1}}$ is a direct edge from $v_t \in V_{\RNum{2}}$ to $v_\circledcirc \in V_{\RNum{3}}$, and $e \in E_{\RNum{2}}$ is a direct edge from $v_v \in V_{\RNum{3}}$ to $v_t \in V_{\RNum{2}}$, labeled by the position of the variable $v_v$ in triple pattern $v_t$, i.e. S, P, or O, . 
\item Weight function $f_w:w \rightarrow  V_{\RNum{2}} $ assigns each vertex $v \in V_{\RNum{2}}$ a weight $w(v)$.
\item Label function $f_l:l \rightarrow  E_{\RNum{2}} $ assigns each edge $e \in E_{\RNum{2}}$ a label $l(e)$, $l(e) \in \{S, O, P\}$.
\end{itemize}
\end{definition}
Intuitively, in a QRG, the subgraph $G_\mathbb{V}=(V_{\RNum{2}} \cup V_{\RNum{2}}, E_{\RNum{1}}, f_l)$ models the \textit{variable correlations} $\mathbb{V}$, and the subgraph  $G_\mathbb{S}=(V_{\RNum{2}} \cup V_{\RNum{3}}, E_{\RNum{2}})$ models the \textit{query semantics} $\mathbb{S}$. The data statistics is introduced by the weight function $f_w$.

The construction of QRG is quite straightforward. Firstly, $G_\mathbb{V}$ can be structured following the operator tree derived from the standard SPARQL query parser (e.g. \textit{Rasqal RDF Query Library}\footnote{\scriptsize{\url{https://github.com/dajobe/rasqal}}}), only adding direction of edges from child vertex to its parent vertex in a bottom-up manner. Then, the estimated cardinality of each triple pattern $T \in \mathbb{T}$ is assigned as a weight to the corresponding vertex $v_t \in V_{\RNum{2}}$. Following, for each unbound variable in $\mathcal{Q}$, a vertex $v_v \in V_{\RNum{3}}$ is added to $G_\mathcal{Q}$. A directed edge from $v_v$ to $v_t \in V_{\RNum{2}}$ which represents the TP $T \in \mathbb{T}$ that contains this variable is established, and labeled with the component position of this variable  (S,O or P) in $T$. 

\begin{figure}
  \centering
  \includegraphics[width=3.2 in]{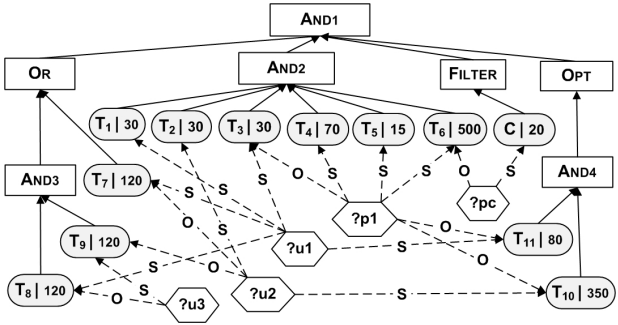}
  \captionsetup{belowskip=0.5pt,aboveskip=0.5pt}
  \caption{Query Relation Graph for $\mathcal{Q}$ in Figure \ref{fig:sparql-query-example}.}
  \label{fig:qrgraph-example}
\end{figure}

\begin{example}
\small
Figure \ref{fig:qrgraph-example} shows a QRG of $\mathcal{Q}_e$ in Figure \ref{fig:sparql-query-example}. It consists of three types of vertices, which are depicted by different kinds of symbols. 
\end{example}

\subsection{Heuristic-Based $\mathcal{CS}$ Planning} \label{subsec-initcs}
Analogous to the query plan selection in CBO, the aim of optimal $\mathcal{CS}$ planning is to select the one with the minimal cost from the search space of all equivalence $\mathcal{CS}$s. This is a classic problem in query optimization, and can be reduced to the problem of finding a minimal weight tree that covers all $v_t \in V_{\RNum{2}}$ in QRG. It is NP-hard as it can be deduced from the well-known TSP problem. Exhaustive exploration is impractical, as $\mathcal{Q}$ may contains tens or even hundreds of TPs. Besides, the techniques borrowed from traditional SQL query optimization may not works well due to lack of schema or integrity constraints. 

Nonetheless, consider that $\mathcal{CS}$ may be dynamically optimized at runtime, we advocate the use of heuristics-based approach to give a reasonable $\mathcal{CS}$ that can be used to guide further optimizations.  To better represent the \textit{query semantics} $\mathbb{S}$ which includes operators like \textsc{Or}, \textsc{Opt} and \textsc{Filter} besides \textsc{And},  we introduce three concepts, namely \textit{Region}, \textit{Ancestor} and \textit{Available Variable}. 

\noindent \textbf{Region}: A \textit{Region} is comprised of all TPs that connects to the same operator $v_{\circledcirc} \in V_{\RNum{1}}$.  The relative position of a \textit{Region} is decided by the position of $v_{\circledcirc} \in V_{\RNum{1}}$ in $G_\mathbb{V}$. More formally, A Region $R(\circledcirc)$ of operator $\circledcirc$ is defined as follows:
\begin{equation*}
\scriptsize
\begin{array}{lr}
R(\circledcirc) = \{v_t | \exists e(v_t \rightarrow v_{\circledcirc}), e \in E_{\RNum{1}}, v_{\circledcirc} \in V_{\RNum{1}}, v_t \in V_{\RNum{2}}\}
\end{array}
\end{equation*}
\normalsize

\noindent \textbf{Ancestor and LCA}: For $v_v \in V_{\RNum{3}}$, the set of its \textit{Ancestor}, $A(v_v)$,  contains all the reachable (denoted as $\rightsquigarrow$)  operators $v_{\circledcirc} \in V_{\RNum{1}}$, and its \textit{Least Common Ancestor} (LCA), $LCA(v_v)$, is an operator $v_{\circledcirc} \in V_{\RNum{1}}$ that is reachable to all $v_{\circledcirc} \in A(v_v)$ with the shortest distance. 
\begin{equation*}
\scriptsize
\begin{array}{lr}
A(v_v) = \{v_\circledcirc | \exists v_v \rightsquigarrow v_{\circledcirc}, v_{\circledcirc} \in V_{\RNum{1}}, v_v \in V_{\RNum{3}} \}; \\ \\
LCA(v_v) = argmin_{\sum d( v_\circledcirc \rightsquigarrow  \hat{v}_\circledcirc)} (\hat{v}_\circledcirc), v_\circledcirc \in A(v_v) , \hat{v}_\circledcirc \in V_{\RNum{1}}.  
\end{array}
\normalsize
\end{equation*} 
where $d( v_\circledcirc \rightsquigarrow  \hat{v}_\circledcirc)$ is the distance of the shortest path from $v_\circledcirc$ to $\hat{v}_\circledcirc$ in $G_\mathcal{Q}$.

\noindent \textbf{Available variable}: For $v_v \in V_{\RNum{3}}$, it is called as an \textit{available variable} in a Region $R(\circledcirc)$, if all its connected $v_t \in V_{\RNum{2}}$ in $R(\circledcirc)$ have been selected as candidates in $\mathcal{CS}$. 

\begin{example}
\small
For $G_\mathcal{Q}$ shown as \textit{Figure} \ref{fig:qrgraph-example}, $T1$-$T6$ are all connected to the $\textsc{And}_2$ operator, they are in a \textit{Region}, $R(\textsc{And}_2)$. Furthermore, according to \textit{Property} \ref{property-exchangable}, $R(\textsc{And}_2)$ is an \textit{Exchangeable Region}. For vertex $?u1 \in  V_{\RNum{3}}$, $A(?u1)=\{\textsc{And}_1$,$\textsc{And}_2$,$\textsc{Or}$, $\textsc{Opt}$,$\textsc{And}_3$,$\textsc{And}_4\}$, and $LCA(?u1)=\textsc{And}_1$. Consider $\widetilde{\mathcal{CS}}=(((T_5$ \textsc{And} $T_4)$ \textsc{And} $(T_6$ \textsc{Filter} C$))$ \textsc{And} $T_3)$, $?p1$ is available in $R(\textsc{And}_2)$, as all its connected $v_t \in V_{\RNum{2}}$ (i.e., $T_4$, $T_5$, $T_6$, $C$) have been arranged in $\widetilde{\mathcal{CS}}$. While for $\widetilde{\mathcal{CS}}$, $?u1$ is not available in $R(\textsc{And}_2)$ until $T_1$ is arranged.
\end{example}

Using the concepts introduced above, we propose the heuristics adopted in $\mathcal{CS}$ planning. The rational behind heuristic-based method is due to that RDF data always have certain fixed patterns among triples that fits the structure of $\mathcal{Q}$. Using these particular features can produce promising results, as stated in \cite{Stocker:2008:SBG:1367497.1367578,Tsialiamanis:2012:HQO:2247596.2247635}.  Based on this, we propose the following heuristics:
\begin{itemize} [topsep=0.5pt,itemsep=-0.5ex,partopsep=1ex,parsep=1ex]
\item \textbf{H1}: The vertices $v \in V_{\RNum{2}}$ is ordered by the labels of theirs incoming edges as: $\varnothing \prec P \prec S \prec O \prec S,P \prec P,O \prec S,O \prec S,P,O$, where $\varnothing$ means no incoming edges. Additionally, for $v \in V_{\RNum{2}}$ with the same types of incoming edges, the less weighted vertex is preferential. This heuristic is a generalization of H1 defined by \cite{Tsialiamanis:2012:HQO:2247596.2247635} on condition that the basic summary data is available.
\item \textbf{H2}: $v_t \in V_{\RNum{2}}$ in the same \textit{Exchangeable Region} and connected to the same $v_v \in V_{\RNum{3}}$ using the same label can be arranged in a successive manner following H1. If $v_v$ has been \textit{Available} in a \textit{Region},then $v_t \in V_{\RNum{2}}$ in other \textit{Region}s and is connected with $v_v$ can be chosen with priority. This heuristic favours the star-shaped subqueries.
\item \textbf{H3}: Evaluates \textsc{Filter} at the earliest possible point. A $v_v \in V_{\RNum{3}}$ that is related to more \textsc{Filter}s has a higher selectivity. 
\item \textbf{H4}: For an $R(\textsc{Opt}$, all TPs can not be planned unless all of the $v_v \in V_{\RNum{3}}$ that these TPs connected with are \textit{available}. This is because that \textsc{Opt} corresponds to \textsc{Left Outer Join}, which is not associative or distributive over other operators.
\end{itemize}

Adopting these heuristics, we propose a greedy $\mathcal{SC}$ generation algorithm, given as \textit{Algorithm} \ref{alg:cs-generation}. The algorithm can be divided into the \textit{Arrange} phase and the \textit{Generation} phase. 

\begin{algorithm}
\caption{\scriptsize{Greedy algorithm for $\mathcal{CS}$ generation.}}
\label{alg:cs-generation}
\scriptsize{}
\SetKwInOut{Input}{input}
\SetKwInOut{Output}{output}
\Input{$G_\mathcal{Q}$, the QRG of query $\mathcal{Q}$.}
\Output{ $\mathcal{CS}$, a candidate sequence. }
\nonl \Begin{
$T^{(1)} \Leftarrow $ chosen according to H1; \label{alg2:func:step0}
$V_{\RNum{2}}^{a} \Leftarrow \{T^{(1)}\}$; $T^c \Leftarrow T^{(1)}$\;
\While {$V_{\RNum{2}}^{a} \subset V_{\RNum{2}} $  } {
$V^{c} \Leftarrow \{(v^c | T^c \subset R(\circledcirc) \wedge v^c \in R(\circledcirc) ) \wedge  (v^c \notin V_{\RNum{2}}^{a})\}$ \; \label{alg2:func:step1}
\If {$V^{c}$ is $\varnothing$} {  \label{alg2:func:step2}
	$V_{\RNum{3}}^{a} \Leftarrow \{v | v \rightarrow v_t, v \in V_{\RNum{3}}, v_t \in V_{\RNum{2}}^{a}\}$ \;
	\For {$\forall v \in V_{\RNum{3}}^{a}$} {
		$V_{\RNum{2}}^{c} \Leftarrow \{ v^c | v \rightarrow v^c, v \in V_{\RNum{3}}^{a}, v^c \in V_{\RNum{2}} \wedge v^c \notin V_{\RNum{2}}^{a} \} $ ;		$V^{c} \Leftarrow V_{\RNum{2}}^{c}$  \;
	}
	$V^{v} \Leftarrow \{ v | v \rightarrow T, \forall T \in V_{\RNum{2}}^{a}  \} $	\;
} \label{alg2:func:step21}
\lElse{
	$V^{v} \Leftarrow \{ v | v \rightarrow T^c \} $ 
}     \label{alg2:func:step3}
\For {$\forall v \in V^{v}$} {  \label{alg2:func:step4}
\If {\textsc{Filter} $\in A(v)$} { \label{alg2:func:step5}
	Arranges all \textsc{Filter} conditions \;
}  \label{alg2:func:step6}

\If {\textsc{Opt} $\in A(v)$ } { \label{alg2:func:step7}
	\If { $(LCA(v)=$ \textsc{Opt}$)\ \vee$ $(V^{c} \subseteq V_{\RNum{2}}^{a})$} {
		\textbf{break} \;
	}
	\lElse{
		\textbf{continue}
	}
}
} \label{alg2:func:step8}

Chosen $T^{(i)}$ from $V^{c}$ using H1,H2 \;  \label{alg2:func:step9}
$V_{\RNum{2}}^{a} \Leftarrow T^{(i)}$; $T^c \Leftarrow T^{(i)}$  \;  
\eIf{$T^c \in R(\circledcirc)$ $\wedge$ $T^{(i)} \in R(\circledcirc)$ } { 
	Appends Region operator $\circledcirc$ and $T^{(i)}$ to $\mathcal{CS}$ \;  \label{alg2:func:step10}
}{
	Checks LCA operator, and adds the operators in the paths from LCA to $T^c$ and $T^{(i)}$ to $\mathcal{CS}$ \;   \label{alg2:func:step11} 
}
}
\scriptsize{}
}
\end{algorithm}

In the \textit{Arrange} phase, it begins with picking the most selective TP $T^{(1)}$ using H1, and $T^{(1)}$ is assigned to $T^c$ as the seed TP to start with (\textit{Line} \ref{alg2:func:step0}). Then, from the \textit{Region} $R(\circledcirc)$ that contains $T^c$, all TPs that are not arranged in $\mathcal{CS}$ are selected in the candidate set, $V^c$ (\textit{Line} \ref{alg2:func:step1}). Meanwhile, the \textit{available variable}s, i.e., $v_v \in V_{\RNum{3}}$ that connected to the arranged TPs, is maintained in a set $V^v$  (\textit{Line} \ref{alg2:func:step3}). If $V^c$ is an empty set, which states that all TPs in $R(\circledcirc)$ have been arranged, the algorithm explores all available $v_v \in V_{\RNum{3}}$, traverses to the \textit{Region}s that contains TPs which are connected with these  \textit{available variable}s. Then all TPs contained in these \textit{Region}s are held in $V^c$ as candidates (\textit{Line} \ref{alg2:func:step2}-\ref{alg2:func:step21}). After determining $V^c$ and  $V^v$ for current $T^c$, the connected \textsc{Filter} operators are checked and arranged if exists using H3 (\textit{Line} \ref{alg2:func:step5}-\ref{alg2:func:step6}). After that, all connected \textsc{Opt} operators are also checked. The arrangement of \textsc{Opt} related TPs is a bit complicated. It is achieved by checking all connected variables in $V^v$. If all of these variables are \textit{available}, or the LCA of these variables are \textsc{Opt}, then all the connected TPs can be arranged using H4. (\textit{Line} \ref{alg2:func:step7}-\ref{alg2:func:step8}). 

In the \textit{Generation} phase, a $\mathcal{CS}$ is incrementally generated by appending the chosen TP $T^{(i)}$ and its related operators to the preceeding $\mathcal{CS}$. If $T^{(i)}$ and $T^c$ are in the same \textit{Region} $R(\circledcirc)$, they can be directly added with the operator $\circledcirc$ using H1 and H2 (\textit{Line} \ref{alg2:func:step10}). If they are from different \textit{Region}s, there a traversal of the operator vertex $v \in V_{\RNum{1}}$ is needed for LCA determination. All the traversed vertices are arranged in $\mathcal{CS}$ accordingly (\textit{Line} \ref{alg2:func:step11}).  It is noteworthy that such LCA-based method is also extensively adopted in traditional query plan generation. The process of \textit{Line} \ref{alg2:func:step1}-\ref{alg2:func:step11} iterates until all TPs and operators have been arranged in a $\mathcal{CS}$.  

\begin{example} \label{example-csgen}
\small
Consider the QRG in \textit{Figure} \ref{fig:qrgraph-example}. $T_5$ is first chosen according to H1. Then the $R(\textsc{And}_2)$ that contains $T_5$ is explored. As $T_4$ and $T_6$ are all connect to vertex \underline{?p1} with the same edge label S, they are chosen accordingly using H2. There $T_6$ also connects to vertex ?\underline{pc}, and \underline{?pc} has a \textsc{Filter} connected. According to H3, compared with $T_3$, $C$ and \textsc{Filter} is arranged with priority. So far, we have got a partial $\widetilde{\mathcal{CS}} = (((T_1\ \textsc{And}\ T_4) \textsc{And} (T_6\ \textsc{Filter}\ C)) \textsc{And}\ T_3)$, with \underline{?p1},\underline{?pc} and \underline{?u1} available. Next, $T_1$ and $T_7$ in $R(\textsc{Or})$ is considered. There according to H4, $R(\textsc{And}_4)$ will not be considered until \underline{?p1}, \underline{?u1} and \underline{?u2} are available in all \textit{Region}s. So $T_{11}$ and $T_{10}$ are arranged lately. Incrementally, we get a final  $\mathcal{CS}$=$((((((T_5$ \textsc{And} $T_4)$ \textsc{And} $(T_6$ \textsc{Filter} C$))$ \textsc{And} $T_3)$ \textsc{And} $T_2)$ \textsc{And} $T_1)$ \textsc{And} $(T_7$ \textsc{Or} $(T_9$ \textsc{And} $T_8)))$ \textsc{Opt} $(T_{11}$ \textsc{And} $T_{10}))$ .
\end{example}

\subsection{$\mathcal{CS}$-to-SQL Conversion}
Intuitively, a $\mathcal{CS}$ represents an operator tree, with TPs as tree leafs, and operators as internal nodes. As stated in \cite{sparql2sql}, the semantics of these operators are analogous between SPARQL and SQL. Specifically, there are direct mapping of \textsc{Or} to an algebraic \textit{Union}, \textsc{Optional} to a \textsc{Left Outer Join}, \textsc{And} to a \textit{Join}, and \textsc{Filter} with constraints can be regarded as a \textit{Selection}. Supporting comprehensive SPARQL specification is a non-trivial task. This needs semantics-preserving SPARQL to relational algebra conversion, and is beyond the scope of this paper. There we focus on the \textsc{Select} query with  \textsc{And}, \textsc{Or}, \textsc{Optional} and \textsc{Filter} operators. Such simplifications are pragmatic and can cover substantially all real-world SPARQL queries, as stated in \cite{Picalausa:2011:RSQ:1999299.1999306}. Besides, there are equivalent specifications in both SPARQL and SQL. i.e. \textsc{Order By}, \textsc{Distinct}, \textsc{Limit}, \textsc{Offset}, \textsc{Not Exists}, and \textsc{Minus}. All these can be natively supported by ROSIE.

%
%
\section{Runtime Optimization} \label{sec:optimization}
Due to the lack of precise statistics, the optimality of a derived $\mathcal{CS}$ can not be guaranteed. The effectiveness and efficacy of such runtime optimization relies on the discrimination of unacceptable biased steps in a $\mathcal{CS}$. In this section, we introduce an optimization technique based on approximate error propagation. 

\subsection{Basic ideas} \label{subsec-rogi}
Generally, based on the definition of cardinality estimation error  (\textit{Equation} \ref{equ-errordef}), the following \textit{Properties} describe the propagation of errors for different operators.
\begin{property} \label{property-propagation}
The cardinality error can be esimated as,
\scriptsize
\begin{enumerate}[topsep=0.5pt,itemsep=-0.5ex,partopsep=1ex,parsep=1ex]
\item $\varepsilon(\mathcal{CS}_j\ \textsc{And}\ \mathcal{CS}_k) = \varepsilon(\mathcal{CS}_j) \times \varepsilon(\mathcal{CS}_k) \times \varepsilon(p(\mathcal{CS}_j\ \textsc{And}\ \mathcal{CS}_k))$;
\item $\varepsilon(\mathcal{CS}_j\ \textsc{Opt}\ \mathcal{CS}_k) = \varepsilon(\mathcal{CS}_j) \times \varepsilon(\mathcal{CS}_k) \times \varepsilon(p(\mathcal{CS}_j\ \textsc{Opt}\ \mathcal{CS}_k))$;
\item $\varepsilon(\mathcal{CS}_j\ \textsc{Or}\ \mathcal{CS}_k) \leq max(\varepsilon(\mathcal{CS}_j) , \varepsilon(\mathcal{CS}_k))$;
\item $\varepsilon(\mathcal{CS}_j\ \textsc{Filter}\ C) = \varepsilon(p(C))$. 
\end{enumerate} 
\normalsize
\end{property}
where $\mathcal{CS}_{i}$ and $\mathcal{CS}_{j}$ are two partial $\mathcal{CS}$, and $C$ is a condition expression for \textsc{Filter}. 

To better illustrate our ideas, we consider a fundamental circumstance that only $\circledcirc = $ \textsc{And} involved. Lets $\mathcal{CS}_{i}$ denotes a $\mathcal{CS}$ at step $i$, and $\widehat{\mathcal{CS}}_{i}$ is one of its equivalence.  We consider three consecutive steps $\mathcal{CS}_{i-1}$, $\mathcal{CS}_{i}$ and $\mathcal{CS}_{i+1}$ that have:
\begin{equation*}
\scriptsize
\begin{array}{lr}
\mathcal{CS}_{i+1} = \mathcal{CS}_{i}\circledcirc T_{i+1} = ((\mathcal{CS}_{i-1} \circledcirc T_i)\circledcirc T_{i+1}); \\
\widehat{\mathcal{CS}}_{i+1} = \widehat{\mathcal{CS}}_{i}\circledcirc T_{i} = ((\mathcal{CS}_{i-1} \circledcirc T_{i+1})\circledcirc T_{i}).
\end{array}
\end{equation*} 
where $T_i$ and $T_{i+1}$ are TPs. As $\mathcal{CS}_{i+1}$ is assumed to be optimal, then the condition  $|\mathcal{CS}_{i+1}|_E \leq |\widehat{\mathcal{CS}}_{i+1}|_E$ holds. As $\circledcirc$ satisfies \textit{Property} \ref{property-exchangable}, then $\mathcal{CS}_{i+1} \rightleftharpoons \widehat{\mathcal{CS}}_{i+1}$ , and we have $|\mathcal{CS}_{i+1}|_P = |\widehat{\mathcal{CS}}_{i+1}|_P$. Lets $\varepsilon_{i+1}$ and $\widehat{\varepsilon}_{i+1}$ denote the estimation error of  $|\mathcal{CS}_{i+1}|$ and $|\widehat{\mathcal{CS}}_{i+1}|$ respectively, we get:
\begin{equation}\label{equ-errorcondition}
\scriptsize
\begin{array}{lr}
\frac{|\mathcal{CS}_{i+1}|_P}{|\mathcal{CS}_{i+1}|_E} \leq \frac{|\widehat{\mathcal{CS}}_{i+1}|_P}{|\widehat{\mathcal{CS}}_{i+1}|_E}  \Rightarrow \ \varepsilon_{i+1} \leq \widehat{\varepsilon}_{i+1}  
\end{array}
\normalsize
\end{equation}
\textit{Equation} \ref{equ-errorcondition}  states that the error is not propagated in a relatively incremental fashion in an optimal $\mathcal{CS}$. If \textit{Equation} \ref{equ-errorcondition} satisfied,  we can say that an optimal $\mathcal{CS}$ is greedily chosen at step $i$. Otherwise, we need to re-consider the next $i$th step by materializing $\mathcal{CS}_{i-1}$. Though the idea is straightforward, actually the error $\varepsilon$ and $\widehat{\varepsilon}$ at step $i$ and $i+1$ are not readily available, as the actual cardinality $|.|_P$ is unknown. In fact, derive the exact cardinalities are costly and in most cases unrealistic. But rely on the available statistics, a theoretical bound of $|.|_P$ can be estimated, which is useful in determining the possible error-prone step in runtime optimization. Next, we explain the method for error bounds estimation by means of $|.|_P$ bounds determination.

\subsection{Error Bound Estimation} \label{subsec-roebe}
Basically, overestimating the error leads to unnecessary materialization steps, while underestimating the error can lead to suboptimal $\mathcal{CS}$. As materialization is a costly operation, and independence assumption tends to underestimate results, it is reasonable to only consider the case of underestimating. As discussed before, in a RDB, the precise cardinality of each distinct elements in $\mathcal{D}$ is generally precomputed and maintained as single-value histogram. We denote these statistics as $|s|$, $|o|$ and $|p|$ for S,P and O respectively. The problems turns to, given $|s|$, $|o|$ and $|p|$, how to estimate a $|\mathcal{CS}|_P$. 

We start with the cardinality estimate error of a single TP. There are seven types of TPs, i.e., (\textit{s p ?o}), (\textit{?s p o}), (\textit{s ?p o}), (\textit{?s p ?o}), (\textit{s ?p ?o}), (\textit{?s ?p o}) and (\textit{?s ?p ?o}). Intuitively, the cardinality of a TP that contains more than one unbound variables can be exactly captured. i.e. $|p|$, $|s|$, $|o|$ and $|\mathcal{D}|$ for (\textit{?s p ?o}), (\textit{s ?p ?o}), (\textit{?s ?p o}) and (\textit{?s ?p ?o}) respectively.  It is a bit complicated for patterns that contain only one unbound variable, as S, P, O can not be assumed as mutually independent. It seems impossible to pre-compute correlations in full generality with reasonable efforts. We resort to estimate a bound for the selectivity, as the correlation varies from a functional dependency to complete independence. To be specific, taken pattern $T \in $ (\textit{?s p o}) for example, $|T|_P= p(?p=p) \cdot p(?o=o|?p=p) \cdot |\mathcal{D} | \in [max(1,\frac{|p|\cdot|o|}{|\mathcal{D}|}), min(|p|,|o|)]$, where the tighter lower bound comes from the independence assumption, and  the upper bound states a functional independence using containment assumption\footnote{\scriptsize{We implicitly assume that a TP $T$ has a least one match, $|T| \geq 1$. }}. This leads to a slight over-estimation, but in practice, it is far more accurate than assuming independence. Similiarily, we can get the bound as follows:
\begin{eqnarray} \label{equ-tperrorbound}
\small
 |T|_P  \in 
 \left\{  
          \begin{array}{lr}  
             \ [max(1,\frac{|p|\cdot|s|}{|\mathcal{D}|}), min(|p|,|s|)], \ T \in (s\ p\ ?o); \\  
             \ [max(1,\frac{|p|\cdot|o|}{|\mathcal{D}|}), min(|p|,|o|)], \ T \in (?s\ p\ o);\\  
             \ [max(1,\frac{|s|\cdot|o|}{|\mathcal{D}|}), min(|s|,|o|)], \ T \in (s\ ?p\ o);\\  
              \end{array}  
\right.
\normalsize
\end{eqnarray}

Next, we introduce operators to TPs. To better explain, we focus on \textsc{Join}, as it shows a multiplicative nature in error propagation. An accurate cardinality estimate of \textsc{Join}s are difficult, as gathering and maintaining the statistics of $\mathcal{D}$ on all possible joinable TPs is an expensive task, even if optimizations such as introducing the semantics of data are considered. Luckily, compared with cardinality estimate, error estimate is a more moderate problem, as it does not needs high accuracy as cardinality estimate does in CBO. Thus it is feasible to derive a reasonable error bound with lower cost. To get a tighter bound, we try to exploit as much information about correlations as possible. Consider the partial $\widetilde{\mathcal{CS}} = ((T_m \ \textsc{And}\  \cdots )\ \textsc{And}\ T_n)$. The cardinality of $\widetilde{\mathcal{CS}}$ can be estimated as:
\begin{equation} \label{equ-cserrorbound}
\scriptsize
|\widetilde{\mathcal{CS}}|_P=(\prod_{i=m}^{n-1}p(T_{i}\ \textsc{And}\ T_{i+1}))\cdot (\prod_{i=m}^{n}|T_i|)
\normalsize
\end{equation}
where $p(T_{i}\ \textsc{And}\ T_{i+1})$ is the selectivity of the \textsc{Join} predicate between $T_i$ and $T_{i+1}$, if it exists. Besides seven type of TPs, there are six types of \textsc{Join}, namely SS-, SO-, OO-, SP-, OP- and PP-\textsc{Join}.  Similarly, the lower and upper bound can be estimated under containment assumption and fully correlated circumstance. That is,  
\begin{eqnarray} \label{equ-joinerrorbound}
\scriptsize
 p(T_{i}\ \textsc{And}\ T_{i+1})  \in 
 \left\{  
          \begin{array}{lr}  
             \ [\frac{1}{|T_i|\cdot|T_{i+1}|},1], \\
             \ \ \ \ \ \ \ \text{\textsc{And} is SO-/SP-/OP-\textsc{Join}}; \\  
             \\
             \ [\frac{1}{|T_i|\cdot|T_{i+1}|}, \frac{1}{max(|T_i|,|T_{i+1}|)}], \\
             \ \ \ \ \ \ \ \text{\textsc{And} is SS-/OO-/PP-\textsc{Join}}.\\  
             \end{array}  
\right.
\normalsize
\end{eqnarray}
The estimation of $\varepsilon(\widetilde{\mathcal{CS}})$ varies according to TP and \textsc{Join} types. Different combination of TP and \textsc{Join} types have different impact on error propagation. This should be considered individually using \textit{Equation} \ref{equ-tperrorbound}, \ref{equ-joinerrorbound} and \ref{equ-cserrorbound}. 

Similarity, we can estimate the error bound following \textit{Property} \ref{property-propagation}, and revise $\mathcal{CS}$ according to \textit{Property} \ref{property-commutative} or \textit{Property} \ref{property-exchangable}. They will not be detailed in this paper.  In some cases this bound is not very useful, as the upper bound can be very large because it assumes fully correlation in data and derives a cardinality of Cartesian product. Without introducing extra statistical data which might change the histogram implementation of the underlying RDB, we will have to fall back to guessing at some point, and scale down the upper bound according to a predefined percentage. This has no theoretical guarantee but can greatly reduce the number of incremental evaluations.

\begin{example}
\scriptsize
 For  $\mathcal{Q}_e$ in Figure \ref{fig:sparql-query-example}, $\mathcal{CS}=((T_4\ \textsc{And}\ T_6) \textsc{And}\ T_3)$, and one of its equivalence $\widehat{\mathcal{CS}}=((T_4\ \textsc{And}\ T_3)\textsc{And}\ T_6)$. The statistics held by RDB are $|\mathcal{D}|=5000$, $|p=creator\_of|=200$, $|p=content|=500$, $|p=type|=300$ and $|o=Post|=70$. The types of TP are (\textit{?s p ?o}) and (\textit{?s p o}), and the \textsc{Join} types includes SS- and SO-\textsc{Join}. According to \textit{Equation} \ref{equ-tperrorbound}, $|T_4|_P \in [1,70]$, $|T_3|_P=200$, and $|T_6|_P=500$ . Applying \textit{Equation} \ref{equ-joinerrorbound}, we get $|\widetilde{\mathcal{CS}}|_P \in [1,7\times 10^6]$, while $|\mathcal{CS}|_P \in [1,3.5\times 10^4]$. As $|\widehat{\mathcal{CS}}|_P =|\mathcal{CS}|_P$ held, we can infer that $\varepsilon(\mathcal{CS}) \geq \varepsilon(\widehat{\mathcal{CS}})$. This states that $\mathcal{CS}$ may not be not optimal.
\end{example}

%
%
\section{Experiments}\label{sec:experiments}
In this section, we empirically evaluated the RDF query performance on two kinds of large-scale benchmark data, as well as real DBpedia dataset. The evaluations were focused on the quantitative metrics, i.e. the query response time. 

\subsection{Environments, Competitors and Datasets} \label{subsec-exp}
\noindent \underline{\textit{\textbf{Environments}}}. All experiments were performed on a machine with Debian 7.4 in 64-bit Linux kernel, two Intel Xeon E5-2640 2.0GHz processors and 64GB RAM. ROSIE was developed in C++, and was compiled using GCC-4.7.1 with -O3 optimization flag. We used \textit{PostgreSQL} 9.4 as RDB back-end, adopted the relational schema proposed by \cite{Bornea:2013:BER:2463676.2463718}. The default DB configuration file is deployed, only tuned the system parameter \textit{shared\_buffers} to a value of 2.5GB, which enabled more shared memory buffers used. 

\vspace{0.5em} \noindent \underline{\textit{\textbf{Datasets}}}. We adopted two different kinds of benchmarks and real DBpedia dataset, as listed in \textit{Table} \ref{tbl:datastats}.

\begin{table}[ht]
\captionsetup{belowskip=0pt,aboveskip=0pt}
\caption{Statistics and characteristics of RDF dataset.}
\label{tbl:datastats}
\scriptsize
\centering
    \begin{tabular}{ |>{\raggedright\arraybackslash}m{0.75in}  ? >{\centering\arraybackslash}m{0.6in} |  >{\centering\arraybackslash}m{0.6in} | >{\centering\arraybackslash}m{0.6in} |}
    \hline
    Dataset   &   \textbf{LUBM}   &   \textbf{SNIB}   & \textbf{DBpedia} \\
    \hline
    \# Triples      &   4,845M   &   1,774M  & 489M   \\
    \hline
    \# Predicates   &     18   &  48  & 60,223 \\
    \hline
   \end{tabular}
\end{table}


\begin{table*}[ht]
\captionsetup{belowskip=0pt,aboveskip=0pt}

\begin{minipage}[b]{\textwidth}
\caption{Canonical LUBM benchmark query response time (in milliseconds).}
\captionsetup[sub]{margin=0pt,skip=0pt, labelfont=bf, labelformat=parens, textfont=normalfont}
\scriptsize
\label{tbl:exp-lubm}
\centering
  $^{\bigstar}$: Wrong in query result size.
  \quad \textbf{N/A}: Query does not ended in 12 hours. \quad \textbf{g.m.} Geometric mean.

\vspace{\abovedisplayskip}

\begin{tabular}{ | >{\centering\arraybackslash}m{0.5in}  ? >{\raggedleft\arraybackslash}m{0.218in} | >{\raggedleft\arraybackslash}m{0.46in} |>{\raggedleft\arraybackslash}m{0.18in} |>{\raggedleft\arraybackslash}m{0.18in} |>{\raggedleft\arraybackslash}m{0.18in} |>{\raggedleft\arraybackslash}m{0.48in} |>{\raggedleft\arraybackslash}m{0.18in} |>{\raggedleft\arraybackslash}m{0.18in} |>{\raggedleft\arraybackslash}m{0.4in} |>{\raggedleft\arraybackslash}m{0.18in} |>{\raggedleft\arraybackslash}m{0.18in} |>{\raggedleft\arraybackslash}m{0.18in} |>{\raggedleft\arraybackslash}m{0.3in} |>{\raggedleft\arraybackslash}m{0.48in} |>{\raggedleft\arraybackslash}m{0.24in} | }
\hline
        & \multicolumn{1}{c|}{L1} & \multicolumn{1}{c|}{L2}  & \multicolumn{1}{c|}{L3} & \multicolumn{1}{c|}{L4} & \multicolumn{1}{c|}{L5} & \multicolumn{1}{c|}{L6} & \multicolumn{1}{c|}{L7} & \multicolumn{1}{c|}{L8} & \multicolumn{1}{c|}{L9} & \multicolumn{1}{c|}{L10} & \multicolumn{1}{c|}{L11} & \multicolumn{1}{c|}{L12} & \multicolumn{1}{c|}{L13} & \multicolumn{1}{c|}{L14}   & \multicolumn{1}{c|}{g.m.} \\[0.2ex]
\hline
\hline    
\textbf{RDF-3X}  &   2   &   $2.3\times10^6$ &   3   &   5   &   7   &   $5.8\times10^6$ &   7   &   401 &   830,944  &   2   &   362  &  14   &   38,988   &   $3.8\times10^6$  & \tiny{718} \\
\hline    
\textbf{TripleBit}  &  {\cellcolor[gray]{0.8}\textbf{0.2}}   &   56,005$^{\bigstar}$   &   {\cellcolor[gray]{0.8}\textbf{0.3}}    &  2   &   {\cellcolor[gray]{0.8}\textbf{3}}   &   $8.1\times10^6$ &   5   &   169$^{\bigstar}$    &   527,032  &   {\cellcolor[gray]{0.8}\textbf{0.5}}   &   {\cellcolor[gray]{0.8}\textbf{3}}   &   {\cellcolor[gray]{0.8}\textbf{4}}   &   639   &   $7.7\times10^6$  & \tiny{147} \\
\hline    
\textbf{Virtuoso7}  &  2   &   {\cellcolor[gray]{0.8}\textbf{42,521}}   &   1    &  {\cellcolor[gray]{0.8}\textbf{3}}   &   4   &   $6.3\times10^6$ &   {\cellcolor[gray]{0.8}\textbf{4} } &   402    &   440,644  &   3   &   3   &   24 & 3,759   &   $4.5\times10^6$  & \tiny{278} \\
\hline    
\textbf{DB2RDF}  &  42   &   92,572   &   41    &  67   &   58   &   {\cellcolor[gray]{0.8}\textbf{142,375}} &   36  &   266    &   162,772  &   25   &   25   &   52 & {\cellcolor[gray]{0.8}\textbf{562}}    &   {\cellcolor[gray]{0.8}\textbf{218,392}}  & \tiny{584} \\
\hline    
\textbf{ROSIE-S}  &  37   &   102,191   &   52    &  56   &   43   &   147,323 &   35  &   {\cellcolor[gray]{0.8}\textbf{192}}    &   {\cellcolor[gray]{0.8}\textbf{142,324}}  &   26   &   62   &   62  &  1,212  &   229,311  & \tiny{638} \\
\hline    
\textbf{ROSIE}  &  37   &   113,721   &   52    &  56   &   43   &   150,521 &   37  &   193    &   153,702  &   26   &   62   &   59 & 1,236   &   228,723  &  \tiny{648} \\
\hline    
\hline    
\#Results    &  4   &   2,528   &   6    &  34   &   719   &   213,817,916 &   67   &   7,790    &   2,703,043  &   4   &   224   &   56 & 95,522   &   162,211,567 &  \\
\hline    
\end{tabular}

\label{subtbl:lubm20480}
\end{minipage}

\vspace{\belowdisplayskip}

\begin{minipage}[b]{\textwidth}
\captionsetup{belowskip=0pt,aboveskip=0pt}
\caption{ Response time of complex queries on LUBM and SNIB dataset (in milliseconds).}
\captionsetup[sub]{margin=0pt,skip=0pt, labelfont=bf, labelformat=parens, textfont=normalfont}
\scriptsize
\label{tbl:exp-complex}
\centering
  $^{\bigstar}$: Wrong in query result size.
	\quad \textbf{N/A}: Query does not ended in 12 hours.
	\quad \textbf{-}: Query does not currently supported.
	\quad \textbf{g.m.} Geometric mean.
\vspace{\abovedisplayskip}
\begin{tabular}{ | >{\centering\arraybackslash}m{0.50in}  ? >{\raggedleft\arraybackslash}m{0.43in} | >{\raggedleft\arraybackslash}m{0.35in} |>{\raggedleft\arraybackslash}m{0.43in} |>{\raggedleft\arraybackslash}m{0.31in} |>{\raggedleft\arraybackslash}m{0.31in} |>{\raggedleft\arraybackslash}m{0.3in}  |>{\raggedleft\arraybackslash}m{0.20in}  ?| >{\raggedleft\arraybackslash}m{0.36in} |>{\raggedleft\arraybackslash}m{0.35in} |>{\raggedleft\arraybackslash}m{0.32in} |>{\raggedleft\arraybackslash}m{0.32in}  |>{\raggedleft\arraybackslash}m{0.36in}  |>{\raggedleft\arraybackslash}m{0.24in} | }
    \hline
    \multirow{2}{*}{} & \multicolumn{7}{ c ?|}{\textbf{LUBM}} & \multicolumn{6}{ c |}{\textbf{SNIB}}  \\
    \cline{2-14}
    \multirow{2}{*}{} &  \multicolumn{1}{c|}{L15} & \multicolumn{1}{c|}{L16}  & \multicolumn{1}{c|}{L17} & \multicolumn{1}{c|}{L15\textsubscript{sel}} & \multicolumn{1}{c|}{L16\textsubscript{sel}} & \multicolumn{1}{c|}{L17\textsubscript{sel}}  & \multicolumn{1}{c?|}{g.m.} & \multicolumn{1}{c|}{S1} & \multicolumn{1}{c|}{S2}  & \multicolumn{1}{c|}{S3} & \multicolumn{1}{c|}{S4} & \multicolumn{1}{c|}{S5} & \multicolumn{1}{c|}{g.m.} \\
\hline    
\hline    
\textbf{RDF-3X}    &  \tiny{$1.2\times10^6$} & N/A & N/A & 3,783 & 14,968 & 92,775  &   \tiny{50,107}  & \tiny{$2.4\times10^6$} & 16,227  & \centering{-} & 8,792 & \tiny{$5.2\times10^6$}  & \tiny{205,416}  \\
\hline    
\textbf{TripleBit}  & \tiny{$1.0\times10^6$$\scriptstyle^{\bigstar}$}   & 456,658$^{\bigstar}$  & \tiny{$1.6\times10^6$$\scriptstyle^{\bigstar}$}  & 4,748$^{\bigstar}$ & 5,890$^{\bigstar}$ & 6,708  & \tiny{71,805}  & 3,234 & 6,076 & \centering{-} & \centering{-} &   124,143$^{\bigstar}$  &  \tiny{13,416}   \\
\hline    
\textbf{Virtuoso7} & $1.4\times10^6$  & 41,013 & 23,626 & {\cellcolor[gray]{0.8}\textbf{83}} & 951 & 444   & \tiny{6,018}    & {\cellcolor[gray]{0.8}\textbf{1,399}} & 16,140  & 164,666 & 1,751  & 23,387  & \tiny{10,877}   \\
\hline    
\textbf{DB2RDF} &  192,662 & 76,391 &  32,442 & 283 & 562 & 342 &  \tiny{5,441} & 3,527 &  12,213 & 57,326 &  {\cellcolor[gray]{0.8}\textbf{1,207}} &  11,812 & \tiny{8,115} \\
\hline    
\textbf{ROSIE-S}   & 522,079  & 452,394 & 627,048 & 3,202 & 4,253 &  6,342  & \tiny{48,359}  & 8,423 &  120,832 &  512,089 & 23,258  &  85,482  & \tiny{63,546} \\
\hline    
\textbf{ROSIE}   &  {\cellcolor[gray]{0.8}\textbf{84,323}} & {\cellcolor[gray]{0.8}\textbf{29,603}} &{\cellcolor[gray]{0.8}\textbf{17,920}}  &  127 &  {\cellcolor[gray]{0.8}\textbf{472}} & {\cellcolor[gray]{0.8}\textbf{383}}    &  \tiny{3,176}  & 8,712 &  {\cellcolor[gray]{0.8}\textbf{4,232}} & {\cellcolor[gray]{0.8}\textbf{21,306}} & 1,411  &   {\cellcolor[gray]{0.8}\textbf{6,923}}  & \tiny{5,907} \\
\hline    
\hline    
\#Results   &   \tiny{17,200,845}   &   1,116    &  21   &   720   &   10 &   3  &  & 27,318  & 54,062 & 258,707 &  323 &  21  &   \\
\hline    
\end{tabular}
\end{minipage}
\end{table*}

\noindent \textbf{LUBM}\footnote{\scriptsize{\url{http://swat.cse.lehigh.edu/projects/lubm/}}} is a widely used benchmark in both academical researches and industrial applications. It provides an university dataset where all components can be generated in a proportional fashion. We generated a large dataset of 20,480 universities. Remind that to support the original LUBM benchmark queries, the standard way is to inference on \textit{subclasses} and \textit{subproperties} schema. We used the inference engine in Virtuoso 7 to generated all inferred triples. 

\noindent \textbf{SNIB}\footnote{\scriptsize{\url{http://www.w3.org/wiki/Social_Network_Intelligence_BenchMark}}} provides an RDF dataset of a Twitter-like social network that includes resources like users, post, reply, tags and comments, etc. The dataset scales according to the number of users. We generated a SNIB dataset of 15,000 users using S3G2\cite{s3g2-generator}. As it models a social network, SNIB data follows a power-law distribution.

\noindent \textbf{DBpedia}\footnote{\scriptsize{Downloaded from \url{http://wiki.dbpedia.org/Downloads2015-04}}} contains information extracted from Wikipedia, and uses RDF as its data description format. Due to DBpedia data spreads on a broad spectrum of topics, its real-world queries are generally sophisticated in nature. There we used its English-only subsets. 

\vspace{0.5em} \noindent \underline{\textit{\textbf{Competitors}}}. There exists a wide choice on the systems that support SPARQL. Among them, four publicly available and competitive ones were chosen. They can be classified as:

\noindent \underline{\textit{Dedicated Systems}}: \textbf{RDF-3X}\footnote{\scriptsize{Available at \url{https://github.com/gh-rdf3x/gh-rdf3x/}}} maintains all possible permutation of \textit{S},\textit{P} and \textit{O} as indexes, and uses B+-tree to facilitate index lookup for single TP. With index-specific query optimization techniques, as well as runtime techniques like \textit{sideways information passing}, RDF-3X remains a competitive state-of-the-art triple store \cite{rdf3x2010}. \textbf{TripleBit}\footnote{\scriptsize{Available at \url{http://grid.hust.edu.cn/triplebit/TripleBit.tar.gz}}} is based a novel data storage that use compact bit-matrices and vertical partitions. Its query optimizer employs a dynamical TP \textsc{Join} order that favors star-shaped sub-queries, and shows by far the best query performance compared with other dedicated systems\cite{triplebit-vldb2013}.

\noindent \underline{\textit{RDB-backed Systems}} : \textbf{Virtuoso 7}\footnote{\scriptsize{Available at \url{https://github.com/openlink/virtuoso-opensource}}} is a commercial-of-the-shelf column store that supports RDF management. It exploits existing relational RDB techniques by adding functionalities to deal with SPARQL, and adopted a "vectorized query execution" technique as outlined in \cite{virtuoso-bitmap}.  \textbf{DB2RDF}\footnote{\scriptsize{Available at \url{https://github.com/Quetzal-RDF/quetzal/}}} proposes an entity-oriented relational schema that of exploits the merits of property table but is more compact in physical storage. Based on that, it provides an effective SPARQL-to-SQL conversion method for RDF over RDB\cite{Bornea:2013:BER:2463676.2463718}. Compared with ROSIE, DB2RDF differs in that it focused on generating optimized SQL which is committed integrally, leaves the burden of all rest works to RDB.

\subsection{Results Analysis}
The queries used are listed in \textit{Appendix}. Each query was executed 11 times in a consecutive manner in each competitor. As the first time of each round was used for cache warm-up, its results was removed from averaging. Thus we reported the query response time as the arithmetic mean of the rest 10 queries, and rounded them to milliseconds. For each experiment, we given the geometric mean of all queries' response times.

To better evaluate the effects of runtime optimization technique, we also implemented a simplified framework for comparison, denoted as \textbf{ROSIE-S}. ROSIE-S generated a static $\mathcal{CS}$ as described in \textit{Section} \ref{sec:initial},  and used this throughout the query execution, without employed the mechanism described in \textit{Section} \ref{sec:optimization}. 

\noindent \textbf{(E-1)} \underline{\textit{Canonical LUBM Benchmark Results.}} \textit{Table} \ref{tbl:exp-lubm} showed the benchmark query performance on LUBM data. As all these benchmark query are relatively simple in that they contain fewer TPs and only \textsc{And} operator involved, runtime optimization mechanism in ROSIE did not worked. Thus ROSIE and ROSIE-S showed approximately the same query performance. Query \textit{L1}, \textit{L3}, \textit{L4}, \textit{L5}, \textit{L7}, \textit{L8}, \textit{L10}, \textit{L11} exhibit the same characteristics in that they all have high-selective TPs in query, Besides, the result sets of these queries are rather small and are independent of the data size. Even though the RDB-backed methods, i.e.,Virtuoso, DB2RDF and ROSIE, showed an inferior performance, consider that all the query response times are within 1 second, it is hard for a user to feel the difference. Query \textit{L2}, \textit{L6}, \textit{L9}, \textit{L13}  and \textit{L14} are characterized by returning a large size of results. The RDB-based methods performed well on these queries. This can be attributed to the efficiency of RDB in handling large amount of disk-based data.

\noindent \textbf{(E-2)} \underline{\textit{Complex Queries on Bechmark Data.}} \textit{Table} \ref{tbl:exp-complex} showed the performance of more complex queries on LUBM and SNIB data. These queries are "complex" in that they typically contains more \textsc{Join}s, range from 6 to 15, with \textsc{Opt},\textsc{Union} and \textsc{Filter} operators involved. Specifically, Query \textit{L15}-\textit{L17} are \textsc{Join}-intensive queries, and ROSIE consistently outperformed others. Query \textit{L15\textsubscript{sel}}-\textit{L17\textsubscript{sel}} are derived by adding a high selective TP to the corresponding Query \textit{L15}-\textit{L17}. These high selective TPs gave explicit hints for the query optimizer, and are helpful in guiding the optimal query plan generation. ROSI generally outperformed on these queries. For SNIB benchmark, it generated a more correlated data than LUBM,  and Query \textit{S1}-\textit{S5} on SNIB adopted more operators. Analogously, ROSIE achieved approximately the best performance. For all queries in E-2, ROSIE exhibited obvious superior query performances compared with  ROISE-E. This demonstrated the effectiveness of the runtime optimization mechanism applied. 

\noindent \textbf{(E-3)} \underline{\textit{Real Queries on DBpeida.}} \textit{Table} \ref{tbl:exp-dbpedia} showed the performance of queries on DBpedia data. In case of real queries \textit{D1}-\textit{D6}, they are characterized by \textsc{Union}-intensive (\textit{D1}), \textit{Filter}-intensive (\textit{D3}) or \textit{Opt}-intensive (\textit{D6}). RDF-3X and TripleBit are more focused on \textsc{And} queries, this renders them incapable for more expressive queries, and they were not employed in this experiment. Consider geometric means, ROSIE achieved approximately 1.5 times faster than DB2RDF, and 3 times faster than Virtuoso. 

\begin{table}[ht]
\captionsetup{belowskip=0pt,aboveskip=0pt}

\caption{Queries response time on DBpedia dataset.}
\centering
(in milliseconds)
\label{tbl:exp-dbpedia}
\scriptsize
\begin{tabular}{ | >{\centering\arraybackslash}m{0.48in}  ? >{\raggedleft\arraybackslash}m{0.22in} | >{\raggedleft\arraybackslash}m{0.25in} |>{\raggedleft\arraybackslash}m{0.25in} |>{\raggedleft\arraybackslash}m{0.21in} |>{\raggedleft\arraybackslash}m{0.23in} |>{\raggedleft\arraybackslash}m{0.25in} |>{\raggedleft\arraybackslash}m{0.17in}  | }
\hline
        & \multicolumn{1}{c|}{D1} & \multicolumn{1}{c|}{D2}  & \multicolumn{1}{c|}{D3} & \multicolumn{1}{c|}{D4} & \multicolumn{1}{c|}{D5} & \multicolumn{1}{c|}{D6}  & \multicolumn{1}{c|}{g.m.}  \\[0.2ex]
\hline
\hline    
\textbf{Virtuoso7}  &  542   &   235   &   3,873  &  342   &   5,243   &  1,121  & \tiny{998}\\
\hline    
\textbf{DB2RDF}  &  72   &   742   &   5,702  &   92  &   2,734   &  462  & \tiny{573} \\
\hline    
\tiny{\textbf{ROSIE-S}}    &  121 &  2,253 &  7,311  &      812  &  6,723 &  1,889  & \tiny{1,655} \\
\hline    
\textbf{ROSIE}    &  {\cellcolor[gray]{0.8}\textbf{62}} &  {\cellcolor[gray]{0.8}\textbf{212}} &  {\cellcolor[gray]{0.8}\textbf{3,102}}  &    {\cellcolor[gray]{0.8}\textbf{84}}  &   {\cellcolor[gray]{0.8}\textbf{1,078}} &    {\cellcolor[gray]{0.8}\textbf{372}}  & \tiny{332}   \\

\hline    
\hline    
\#Results    &  1,182   &   11   &  58,761  &  53   &   48   &  2   &  \\
\hline    
\end{tabular}
\end{table}

\noindent \textbf{Results Analysis:}  To summarize, E-1 demonstrated the workability of ROSIE framework. The comparative performance of ROSIE with others showed the efficiency of the $\mathcal{CS}$ generated by the heuristic-based planning algorithm, and the results size showed the correctness of query execution in ROSIE.  As shown in E-2 and E-3, for more complex queries on data with more sophisticated correlations, RDB-based systems outperformed the dedicated systems. ROSIE performed much better, achieved 1.5X to 20X better query performance. In addition, compared with ROSIE-S that used a static and heuristic-based query planning, the performance of ROSIE outperformed by orders of magnitudes. These shown that for complicated query and sophisticated data, the essence of ROSIE runtime optimization gets highlighted. We trust that with further optimizations applied for the back-end RDB, ROSIE still have room for boosting its query performance.

%
%

\section{Conclusions} \label{sec:conclusions}

In this paper, we introduced ROSIE (\underline{R}untime \underline{O}ptimization of \underline{S}PARQL query using \underline{I}ncremental \underline{E}valuation), a framework aiming at promoting the query performance, as well as supporting more expressive SPARQL queries. To achieve these, we proposed a heuristic-based approach for TP execution order determination. Furthermore, we devised a mechanism that adopted cardinality estimate error bound determination to optimize the generated order dynamically at runtime. We showed the correctness and effectiveness of ROSIE through extensive experiments on synthetic and real RDF data. There are still many open issues left for future work. Among them, supporting queries that involved more complete SPARQL specifications, and the semantic-preserving SPARQL-to-SQL conversion are the most promising aspects to be explored.

\section*{Acknowledgments}
<Intentionally Left Blank>

%

%
%

%
%
%
%

\section{Appendix}\label{sec-appendix}
\begin{scriptsize}

\subsection{LUBM Queries} \label{appendix-lubmquery}

\vspace{0.5em} \noindent PREFIX rdf: $\langle$http://www.w3.org/1999/02/22-rdf-syntax-ns\#$\rangle$

\vspace{0em} \noindent PREFIX rdfs: $\langle$http://www.w3.org/2000/01/rdf-schema\#$\rangle$

\vspace{0em} \noindent PREFIX ub: $\langle$http://swat.cse.lehigh.edu/onto/univ-bench.owl\#$\rangle$

\vspace{0em} \noindent \textbf{L1-L14:} Same as LUBM Q1-Q14 at \url{http://swat.cse.lehigh.edu/projects/lubm/queries-sparql.txt} respectively.

\vspace{0em} \noindent \textbf{L15:}SELECT ?a1 ?a2 ?a3 ?a4 WHERE\{?a4 rdf:type ub:University. ?a1 rdf:type ub:GraduateStudent. ?a1 ub:advisor ?a2. ?a2 ub:worksFor ?a3. ?a3 ub:subOrganizationOf ?a4. ?a2 rdf:type ub:FullProfessor. ?a3 rdf:type ub:Department. \}

\vspace{0em} \noindent \textbf{L15\textsubscript{sel}:} \textbf{L15} with pattern \{?a4 ub:name "University7". \}

\vspace{0em} \noindent \textbf{L16:}SELECT ?x ?y ?z WHERE\{?x rdf:type ub:GraduateStudent. ?y rdf:type ub:University.
 ?z rdf:type ub:Department.
 ?c rdf:type ub:GraduateCourse.
 ?x ub:takesCourse ?c.
 ?p rdf:type ub:FullProfessor.
 ?p ub:teacherOf ?c.
 ?p ub:worksFor ?z.
 ?x ub:memberOf ?z.
 ?x ub:undergraduateDegreeFrom ?y.
 ?z ub:subOrganizationOf ?y\}

\vspace{0em} \noindent \textbf{L16\textsubscript{sel}:}\textbf{L16} with pattern \{?y ub:name "University6". \}

\vspace{0em} \noindent \textbf{L17:}SELECT ?x ?y ?p ?b ?c WHERE\{?x rdf:type ub:GraduateStudent.
 ?x rdf:type ub:TeachingAssistant.
 ?x ub:advisor ?p.
 ?x ub:takesCourse ?c.
 ?x ub:memberOf ?z.
 ?x ub:undergraduateDegreeFrom ?y.
 ?y rdf:type ub:University.
 ?c rdf:type ub:GraduateCourse.
 ?p ub:worksFor ?z.
 ?p rdf:type ub:AssociateProfessor.
 ?z ub:subOrganizationOf ?y.
 ?z rdf:type ub:Department.
 ?b rdf:type ub:Publication.
 ?p ub:teacherOf ?c.
 ?b ub:publicationAuthor ?x. \}

\vspace{0em} \noindent \textbf{L17\textsubscript{sel}:} \textbf{L17} with pattern \{?Y ub:name "University786" \}.



\subsection{SNIB Queries} \label{appendix-snibquery}

\vspace{0.5em} \noindent PREFIX foaf: $\langle$http://xmlns.com/foaf/0.1/$\rangle$

\vspace{0em} \noindent PREFIX dc: $\langle$http://purl.org/dc/elements/1.1/$\rangle$

\vspace{0em} \noindent PREFIX sioc: $\langle$http://rdfs.org/sioc/ns\#$\rangle$

\vspace{0em} \noindent PREFIX sioct: $\langle$http://rdfs.org/sioc/type\#$\rangle$

\vspace{0em} \noindent PREFIX sib: $\langle$http://www.ins.cwi.nl/sib/vocabulary/$\rangle$


\vspace{0em} \noindent PREFIX dbp: $\langle$http://dbpedia.org/resource/$\rangle$

\vspace{0em} \noindent \textbf{S1:}SELECT ?user ?commentcontent ?commentdate WHERE\{?user foaf:knows ?friend.
 ?user rdf:type sib:User.
 ?friend rdf:type sib:User.
 ?friend sioc:moderator\_of ?forum.
 ?post sib:hashtag dbp:Creek .
 ?post rdf:type sioc:Post.
 ?forum sioc:container\_of ?post.
 ?post sioc:content ?postcontent.
 ?post sioc:container\_of ?postcomment.
 ?postcomment rdf:type sioc:Item.
 ?postcomment sioc:content ?commentcontent.
 ?postcomment dc:created ?commentdate. \}

\vspace{0em} \noindent \textbf{S2:}SELECT ?user1 ?user2 ?friend WHERE\{?user1 foaf:knows ?user2.
 ?user2 foaf:knows ?friend .
 ?friend foaf:knows ?user1 .
 ?user1 rdf:type sib:User .
 ?user2 rdf:type sib:User .
 ?friend rdf:type sib:User .
 ?friend sioc:creator\_of ?post.
 ?post rdf:type sioc:Post .
 ?post sioc:content ?postcontent.
 ?post sib:hashtag dbp:Creek .
 ?post sib:liked\_by ?user1\}


\vspace{0em} \noindent \textbf{S3:} SELECT ?user0 ?friend ?photo WHERE \{
        ?user0 rdfs:type sib:User.
        ?friend rdfs:type sib:User .
        ?photo rdfs:type sib:Photo .
        \{
            ?friend foaf:knows ?user0.
        \}
        OPTIONAL\{
            ?friend foaf:knows ?u2 .
            ?u2 foaf:knows ?user0 .
        \}
        ?photo sib:usertag ?friend.
        ?photo dbp:location "Germany" .
        ?user0 sib:liked\_by ?photo .
        FILTER EXISTS\{
            ?pa sioc:container\_of ?photo .
            ?pa rdfs:type type:ImageGallery .
            \{ ?u2 sioc:creator\_of ?pa . \} UNION \{ ?user0 sioc:creator\_of ?pa . \}
        \}
 \}

\vspace{0em} \noindent \textbf{S4:}SELECT ?user1 WHERE\{
 ?user1 rdf:type sib:User .
 ?user2 rdf:type sib:User .
 ?photo rdfs:type sib:Photo.
 ?user2 sioc:creator\_of ?pa.
 ?photo sib:usertag ?user1.
 ?pa rdfs:type sioct:ImageGallery.
        ?user1 sib:In\_relationship\_with ?user2.
        ?pa sioc:container\_of ?photo.
        ?user2 sioc:creator\_of ?pa.
OPTIONAL \{
?post sib:liked\_by ?user1.
?post sioc:content ?postcontent.
?post sib:hashtag dbp:Creek .
\} \}

\vspace{0em} \noindent \textbf{S5:}SELECT ?u1 ?u2 WHERE\{?f sib:memb ?u1.
 ?f sib:memb ?u2.
 ?u2 foaf:knows ?u1.
 ?u3 foaf:knows ?u2.
 ?u3 foaf:knows ?u1.
 ?u1 rdf:type sib:User.
 ?u2 rdf:type sib:User.
 ?u3 rdf:type sib:User.
 ?p sib:hashtag dbp:Hummelshof.
 ?u1 sioc:creator\_of ?p.
 ?p rdf:type sioc:Post.
 ?forum sioc:container\_of ?p.
 ?forum sioc:container\_of ?c.
 ?c rdf:type sioc:Item.
 ?u2 sioc:creator\_of ?c.
 ?c sioc:reply\_of ?p.
 ?p sib:liked\_by ?u2. \}


\subsection{DBpedia Queries} \label{appendix-snibquery}

\vspace{0.5em} \noindent PREFIX dbp: $\langle$http://dbpedia.org/$\rangle$ 

\vspace{0.em} \noindent PREFIX r: $\langle$http://dbpedia.org/resource/$\rangle$

\vspace{0em} \noindent PREFIX p: $\langle$http://dbpedia.org/property/$\rangle$ 

\vspace{0em} \noindent PREFIX o: $\langle$http://dbpedia.org/ontology/$\rangle$

\vspace{0em} \noindent PREFIX owl: $\langle$http://www.w3.org/2002/07/owl\#$\rangle$ 

\vspace{0em} \noindent PREFIX xsd: $\langle$http://www.w3.org/2001/XMLSchema\#$\rangle$ 

\vspace{0em} \noindent PREFIX skos: $\langle$http://www.w3.org/2004/02/skos/core\#$\rangle$

\vspace{0em} \noindent \textbf{D1:}SELECT ?club WHERE \{ 
     \{ ?club  p:league r:Premier\_League.\}
     UNION
     \{?club p:league  r:Scottish\_Premier\_League.\}
      UNION
     \{?club p:league  r:Football\_League\_One.\}
     UNION
     \{?club p:league  r:Football \_League \_Championship.\}
     UNION
     \{?club p:league  r:Football\_League \_Two.\}
  \}

\vspace{0em} \noindent \textbf{D2:}SELECT DISTINCT ?v1 ?v2 WHERE \{  
?v1 rdf:type o:Wrestler . 
?v2 rdfs:label ?v2 . 
FILTER regex(str(?v2), "sep", "i") 
\} 

\vspace{0em} \noindent \textbf{D3:}SELECT ?v1 ?v2 WHERE \{ 
\{ ?v1 rdf:type o:Settlement.
?v1 p:population ?v2. 
FILTER ( xsd:integer(?v2) > 54 ) \} 
UNION \{ 
?v1 rdf:type o:Settlement.
?v1 p:populationUrban ?v2.
 FILTER ( xsd:integer(?v2) > 54 ) 
\} \}

\vspace{0em} \noindent \textbf{D4:}SELECT ?v1 ?v2 ?v3 ?v4 ?v5 WHERE \{ 
\{ r:Treehouse\_of\_Horror \_XX ?v1 ?v2. 
?v2 foaf:name ?v4. \} 
UNION \{ 
?v3 ?v1 r:Treehouse\_of \_Horror \_XX ; 
?v3 foaf:name ?v5.
 \} \}

\vspace{0em} \noindent \textbf{D5:}SELECT ?v1 ?v2 ?v3 ?v4 ?v5 WHERE\{
?v1 rdf:type o:Artist. ?v1 p:name ?v2. ?v1 p:pages ?v3. ?v1 p:isbn ?v4.  ?v1 p:author ?v5.\} 

\vspace{0em} \noindent \textbf{D6:}SELECT ?V1 ?v2 ?v3 WHERE \{  
?v1 rdf:type o:Settlement.
?v1 rdfs:label "Djanet". 
?v2 rdf:type o:Airport. 
\{ ?v2 o:city ?v1.\} 
UNION \{ ?v2 o:iataLocationIdentifier ?v3. \} 
UNION \{ 
?v2 o:location ?v1. \} 
\{ ?v2 p:iata ?v3. \} 
OPTIONAL \{ ?v2 foaf:homepage ?v5. \} 
OPTIONAL \{ ?v2 p:nativename ?v4. \}
 \} 

\end{scriptsize}

\end{document}